\documentclass[english]{article}
\usepackage[T1]{fontenc}
\usepackage[latin9]{inputenc}
\usepackage{geometry}
\usepackage{color}
\usepackage{babel}
\usepackage{float}
\usepackage{textcomp}
\usepackage{amsmath}
\usepackage{amsthm}
\usepackage{graphicx}
\usepackage{setspace}
\usepackage{esint}
\usepackage[unicode=true,pdfusetitle,
 bookmarks=true,bookmarksnumbered=false,bookmarksopen=false,
 breaklinks=false,pdfborder={0 0 1},backref=false,colorlinks=false]
 {hyperref}

\makeatletter
  \theoremstyle{plain}
  \newtheorem{conjecture}{\protect\conjecturename}
  \theoremstyle{definition}
  \newtheorem{defn}{\protect\definitionname}
  \theoremstyle{plain}
  \newtheorem{lem}{\protect\lemmaname}

\@ifundefined{showcaptionsetup}{}{%
 \PassOptionsToPackage{caption=false}{subfig}}
\usepackage{subfig}
\makeatother

  \providecommand{\conjecturename}{Conjecture}
  \providecommand{\definitionname}{Definition}
  \providecommand{\lemmaname}{Lemma}

\begin{document}
\begin{doublespace}
\begin{center}
\textbf{\textcolor{black}{\large{}Microstructure under the Microscope:
Tools to Survive and Thrive in The Age of (Too Much) Information}}
\par\end{center}{\large \par}

\begin{center}
\textbf{Ravi Kashyap }
\par\end{center}

\begin{center}
\textbf{IHS Markit / City University of Hong Kong }
\par\end{center}

\begin{center}
\begin{center}
\today
\par\end{center}
\par\end{center}

\begin{center}
Microstructure, Marketstructure, Microscope, Dimension Reduction,
Distance Measure, Covariance, Distribution, Uncertainty
\par\end{center}

\begin{center}
\par\end{center}
\end{doublespace}

\begin{center}
\textbf{\textcolor{blue}{\href{http://www.iijournals.com/doi/abs/10.3905/jot.2017.12.2.005}{Edited Version: Kashyap, R. (2017). Microstructure under the Microscope: Tools to Survive and Thrive in The Age of (Too Much) Information. The Journal of Trading, 12(2), 5-27. }}}\tableofcontents{}
\par\end{center}

\begin{doublespace}

\section{Abstract }
\end{doublespace}

\begin{doublespace}
Market Microstructure is the investigation of the process and protocols
that govern the exchange of assets with the objective of reducing
frictions that can impede the transfer. In financial markets, where
there is an abundance of recorded information, this translates to
the study of the dynamic relationships between observed variables,
such as price, volume and spread, and hidden constituents, such as
transaction costs and volatility, that hold sway over the efficient
functioning of the system.

\textquotedblleft My dear, here we must process as much data as we
can, just to stay in business. And if you wish to make a profit you
must process atleast twice as much data.\textquotedblright{} - Red
Queen to Alice in Hedge-Fund-Land.

Necessity is the mother of all invention / creation / innovation,
but the often forgotten father is frustration. In this age of (Too
Much) Information, it is imperative to uncover nuggets of knowledge
(signal) from buckets of nonsense (noise).

To aid in this effort to extract meaning from chaos and to gain a
better understanding of the relationships between financial variables,
we summarize the application of the theoretical results from (Kashyap
2016b) to microstructure studies. The central concept rests on a novel
methodology based on the marriage between the Bhattacharyya distance,
a measure of similarity across distributions, and the Johnson Lindenstrauss
Lemma, a technique for dimension reduction, providing us with a simple
yet powerful tool that allows comparisons between data-sets representing
any two distributions. 

We provide an empirical illustration using prices, volumes and volatilities
across seven countries and three different continents. The degree
to which different markets or sub groups of securities have different
measures of their corresponding distributions tells us the extent
to which they are different. This can aid investors looking for diversification
or looking for more of the same thing.

In Indian mythology, it is believed that in each era, God takes on
an avatar or reincarnation to fight the main source of evil in that
epoch and to restore the balance between good and bad. In this age
of too much information and complexity, perhaps the supreme being
needs to be born as a data scientist, conceivably with an apt nickname,
the Infoman. Until higher powers intervene and provide the ultimate
solution to completely eliminate information overload, we have to
make do with marginal methods, such as this composition, to reduce
information.

As we wait for the perfect solution, it is worth meditating upon what
superior beings would do when faced with a complex situation, such
as the one we are in. It is said that the Universe is but the Brahma's
(Creator's) dream. Research (Effort / Struggle) can help us understand
this world; Sleep (Ease / Peace of Mind) can help us create our own
world. A lesson from close by and down under: We need to \textquotedblleft Do
Some Yoga and Sleep Like A Koala\textquotedblright .
\end{doublespace}

\begin{doublespace}

\section{Objectively Subjective}
\end{doublespace}

\begin{doublespace}
A hall mark of the social sciences is the lack of objectivity. Here
we assert that objectivity is with respect to comparisons done by
different participants and that a comparison is a precursor to a decision.
\end{doublespace}
\begin{conjecture}
\begin{doublespace}
\textbf{Despite the several advances in the social sciences,} \textbf{we
have yet to discover an objective measuring stick for comparison,
a so called, True Comparison Theory, which can be an aid for arriving
at objective decisions}. \end{doublespace}

\end{conjecture}
\begin{doublespace}
The search for such a theory could again be compared, to the medieval
alchemists\textquoteright{} obsession with turning everything into
gold (Kashyap 2014a). For our present purposes, the lack of such an
objective measure means that the difference in comparisons, as assessed
by different participants, can effect different decisions under the
same set of circumstances. Hence, despite all the uncertainty in the
social sciences, the one thing we can be almost certain about is the
subjectivity in all decision making.
\end{doublespace}

\subsection{Merry-Go-Round of Comparisons, Decisions and Actions}

\begin{doublespace}
This lack of an objective measure for comparisons, makes people react
at varying degrees and at varying speeds, as they make their subjective
decisions. A decision gives rise to an action and subjectivity in
the comparison means differing decisions and hence unpredictable actions.
This inability to make consistent predictions in the social sciences
explains the growing trend towards comprehending better and deciphering
the decision process and the subsequent actions, by collecting more
information across the entire cycle of comparisons, decisions and
actions. Another feature of the social sciences is that the actions
of participants affects the state of the system, effecting a state
transfer which perpetuates another merry-go-round of comparisons,
decisions and actions from the participants involved. This means,
more the participants, more the changes to the system, more the actions
and more the information that is generated to be gathered.

Restricted to the particular sub-universe of economic and financial
theory, this translates to the lack of an objective measuring stick
of value, a so called, True Value Theory. This lack of an objective
measure of value, (hereafter, value will be synonymously referred
to as the price of a financial instrument), makes prices react at
differing degrees and at varying velocities to the pull of different
macro and micro factors.

(Lawson 1985) argues that the Keynesian view on uncertainty (that
it is generally impossible, even in probabilistic terms, to evaluate
the future outcomes of all possible current actions; Keynes 1937;
1971; 1973), far from being innocuous or destructive of economic analysis
in general, can give rise to research programs incorporating, amongst
other things, a view of rational behavior under uncertainty, which
could be potentially fruitful. (McManus and Hastings 2005) clarify
the wide range of uncertainties that affect complex engineering systems
and present a framework to understand the risks (and opportunities)
they create and the strategies system designers can use to mitigate
or take advantage of them. These viewpoints hold many lessons for
policy designers in the social sciences and could be instructive for
researchers looking to create methods to compare complex systems,
keeping in mind the caveats of dynamic social systems.
\end{doublespace}

\begin{doublespace}

\subsection{Interplay of Information and Intelligence}
\end{doublespace}

\begin{doublespace}
On the surface, it would seem that there is a repetitive nature to
portfolio management, which we can term \textquotedbl{}The Circle
of Investment\textquotedbl{} (Kashyap 2014b), making it highly amenable
to automation. But we need to remind ourselves that the reiterations
happen under the purview of a special kind of uncertainty that applies
to the social sciences. (Kashyap 2014a) goes into greater depth on
how the accuracy of predictions and the popularity of generalizations
might be inversely related in the social sciences. In the practice
of investment management and also to aid other business decisions,
more data sources are being created, collected and used along with
increasing automation and artificial intelligence.

If Alice and Red Queen of the Wonderland fame (Carroll 1865; 1871;
End-note \ref{enu:The-Red-Queen's}) were to visit Hedge-Fund-Land
(or even Business-Land), the following modification of their popular
conversation would aptly describe the situation today, ``My dear,
here we must process as much data as we can, just to stay in business.
And if you wish to make a profit you must process atleast twice as
much data.''

We could also apply this to HFT-Land and say: \textquotedblleft My
dear, here we must trade as fast as we can, just to stay in business.
And if you wish to make a profit, you must trade atleast twice as
fast as that.\textquotedblright , while reminiscing that the jury
is still out on whether HFT is Good, Bad or Just Ugly and Unimportant.

In Academic-Land, this would become: \textquotedblleft My dear, here
we must process as much data (and include as many strange symbols
or obfuscating terms) as we can, just to create a working paper. And
if you wish to make a publication you must process atleast twice as
much data (and include atleast twice as many strange characters or
obfuscating expressions).\textquotedblright{}

We currently lack a proper understanding of how, in some instances,
our brains (or minds; and right now it seems we don't know the difference!)
make the leap of learning from information to knowledge to wisdom
(See Mill 1829; Mazur 2015 for more about learning and behavior).
The problem of creating artificial intelligence can be a child's play,
depending on which adult's brainpower acts as our gold standard. Perhaps,
the real challenge is to replicate the curiosity and learning an infant
displays. Intellect might be a byproduct of Inquisitiveness, demonstrating
another instance of an unintended yet welcome consequence (Kashyap
2016e). This brings up the question of Art and Science in the practice
of asset management (and everything else in life?); which are more
related than we probably realize, ``Art is Science that we don't
know about; Science is Art restricted to a set of symbols governed
by a growing number of rules'' (Kashyap 2014a).

While the similarities between art and science, should give us hope;
we need to face the realities of the situation. Right now, arguably,
in most cases, we (including computers and intelligent machines?)
can barely make the jump from the information to the knowledge stage;
even with the use of cutting / (bleeding?) edge technology and tools.
This exemplifies three things: 
\end{doublespace}
\begin{enumerate}
\begin{doublespace}
\item We are still in the information age. As another route to establishing
this, consider this: Information is Hidden; Knowledge is Exchanged
or Bartered; Wisdom is Dispersed. Surely we are still in the Information
Age since a disproportionate amount of our actions are geared towards
accumulating unique data-sets for the sole benefits of the accumulators. 
\item Automating the movement to a higher level of learning, which is necessary
for dealing with certain doses of uncertainty, is still far away. 
\item Some of us missed the memo that the best of humanity are actually
robots in disguise, living amongst us.\end{doublespace}

\end{enumerate}
\begin{doublespace}
Hence, it is not Manager versus Machine (Portfolio Manager vs Computing
Machine or MAN vs MAC, in short; End-notes \ref{enu:Portfolio Manager},
\ref{enu:Universal Computing Machine}, \ref{enu:Computer}). Not
even MAN and MAC against the MPC (Microsoft Personal Computer; End-notes
\ref{enu: Mac or Macintosh}, \ref{enu:Personal Computer}, \ref{enu:MAC vs MPC})?
It is MAN, MAC and the MPC against increasing complexity! (Also in
scope are other computing platforms from the past, present and the
future: Williams 1997; Ifrah, Harding, Bellos and Wood 2000; Ceruzzi
2003; End-notes \ref{enu:History Computing}, \ref{enu:Computing Platform},
\ref{enu:Cloud Computing}, \ref{enu:Quantum Computing}). This increasing
complexity and information explosion is perhaps due to the increasing
number of complex actions perpetrated by the actors that comprise
the financial system. The human mind will be obsolete if machines
can fully manage assets and we would have bigger problems on our hands
than who is managing our money. We need, and will continue to need,
massive computing power to mostly separate the signal from the noise.
\end{doublespace}

\begin{doublespace}

\subsection{Simply Too Complex}
\end{doublespace}

\begin{doublespace}
(Simon 1962) points out that any attempt to seek properties common
to many sorts of complex systems (physical, biological or social),
would lead to a theory of hierarchy since a large proportion of complex
systems observed in nature exhibit hierarchic structure; that a complex
system is composed of subsystems that, in turn, have their own subsystems,
and so on. This might hold a clue to the miracle that our minds perform;
abstracting away from the dots that make up a picture, to fully visualizing
the image, that seems far removed from the pieces that give form and
meaning to it. To helps us gain a better understanding of the relationships
between financial variables, we construct a metric built from smaller
parts, but gives optimal benefits when seen from a higher level. Contrary
to what conventional big picture conversations suggest, as the spectator
steps back and the distance from the picture increases, the image
becomes smaller yet clearer.

As a first step, we recognize that one possible categorization (Kashyap
2016c) of different fields can be done by the set of questions a particular
field attempts to answer. The answers to the questions posed by any
domain can come from anywhere or from phenomenon studied under a combination
of many other disciplines. Hence, the answers to the questions posed
under the realm of economics and finance can come from seemingly diverse
subjects, such as, physics, biology, mathematics, chemistry, and so
on. As we embark on the journey to apply the knowledge from other
fields to finance, we need to be aware that finance is Simply Too
Complex, since all of finance, through time, has involved three simple
outcomes - \textquotedblleft Buy, Sell or Hold\textquotedblright .
The complications are mainly to get to these results.
\end{doublespace}
\begin{defn}
\begin{doublespace}
\textbf{\textit{Market Microstructure is the investigation of the
process and protocols that govern the exchange of assets with the
objective of reducing frictions that can impede the transfer.}} 

In financial markets, where there is an abundance of recorded information,
this translates to the study of the dynamic relationships between
observed variables, such as price, volume and spread, and hidden constituents,
such as transaction costs and volatility, that hold sway over the
efficient functioning of the system (Kashyap 2015b).\end{doublespace}

\end{defn}
\begin{doublespace}
While it might be possible to observe historical trends (or other
attributes) and make comparisons across fewer number of entities,
in large systems where there are numerous components or contributing
elements, this can be a daunting task. If time travel were to become
possible, time series would no longer be relevant. We are accustomed
to using time and money as our units of measurement. Time and money
are but means to an end. If we start viewing efforts and the world
in terms of what we hope to accomplish ultimately, it might lead to
better results. 

In this present paper, we put aside the fundamental question of whether
we need complicated models or merely better morals and present quantitative
measures across aggregations of smaller elements that can aid decision
makers by providing simple yet powerful metrics to compare groups
of entities. The results draw upon sources from statistics, probability,
economics / finance, communication systems, pattern recognition and
information theory; becoming one example of how elements of different
fields can be combined to provide answers to the questions raised
by a particular field. The degree to which different markets or sub
groups of securities have different measures of their corresponding
distributions tells us the extent to which they are different. This
can aid investors looking for diversification or looking for more
of the same thing.
\end{doublespace}

\begin{doublespace}

\subsection{Nuggets of Knowledge from Buckets of Nonsense}
\end{doublespace}

\begin{doublespace}
Necessity is the mother of all invention / creation / innovation,
but the often forgotten father is frustration. In this age of (Too
Much) Information, it is imperative to uncover nuggets of knowledge
from buckets of nonsense.

To aid in this effort to extract meaning from chaos, we summarize
the application of the theoretical results from (Kashyap 2016b) to
microstructure studies. The central concept rests on a novel methodology
based on the marriage between the Bhattacharyya distance, a measure
of similarity across distributions, and the Johnson Lindenstrauss
Lemma, a technique for dimension reduction, providing us with a simple
yet powerful tool that allows comparisons between data-sets representing
any two distributions, perhaps also becoming, to our limited knowledge,
an example of perfect matrimony.

We return to Sergei Bubka, our Icon of Uncertainty (Kashyap 2016a).
As a refresher for the younger generation, he broke the pole vault
world record 35 times. We can think of regulatory change or the utilization
of newer methods and techniques as raising the bar. Each time the
bar is raised, the spirit of Sergei Bubka, in all of us, will find
a way over it. The varying behavior of participants in a social system
will give rise to unintended consequences (Kashyap 2016e) and as long
as participants are free to observe the results and modify their actions,
this effect will persist. (Kashyap 2015a) consider ways to reduce
the complexity of social systems, which could be one way to mitigate
the effect of unintended outcomes. While attempts at designing less
complex systems are worthy endeavors, reduced complexity might be
hard to accomplish in certain instances and despite successfully reducing
complexity, alternate techniques at dealing with uncertainty are commendable
complementary pursuits (Kashyap 2016d).

Asset price bubbles are seductive but scary when they burst. What
we learn from the story of Beauty and the Beast is that they must
coexist; we need to learn to love the beast before we can uncover
the beauty. Similarly bubbles and busts must be close to one another.
If we find that microstructure variables, especially implicit trading
costs, are showing steady movement, the change in transaction costs
could be a signal of a potential building up of a bubble and a later
bust. Our study will allow the comparison of trading costs across
aggregations of individual securities, allowing inferences to be drawn
across sectors or markets, enabling us to find early indications of
bubbles building up in corners of the economy.
\end{doublespace}

\begin{doublespace}

\subsection{The Miracle of Mathematics}
\end{doublespace}

\begin{doublespace}
Lastly on a cautionary note, since the concepts mentioned below involve
non-trivial mathematical principles, we point out that the source
of most (all) human conflict (and misunderstanding) is not because
of what is said (written) and heard (read), but is partly due to how
something is said and mostly because of the difference between what
is said and heard and what is meant and understood. We list a few
different ways of describing what mathematics is and perhaps why it
is miraculously magical most of the time but minutiae some times,
that could be relegated to an appendix to be safely ignored.
\end{doublespace}
\begin{enumerate}
\begin{doublespace}
\item Mathematics is built on one simple operation, addition, making it
a fractal with addition as its starting point.
\item Mathematics has become complex because of the confusion that different
notations, assumptions not made explicit and missed steps can create. 
\item Mathematics without the steps is like a treasure hunt without the
clues.
\item Mathematics is like a swimsuit model wearing a Burkha; we need to
see beyond the symbols and the surface to appreciate the beauty.\end{doublespace}

\end{enumerate}
\begin{doublespace}
In a complex system, deriving equations can be a daunting exercise,
and not to mention, of limited practical validity. Hence, to supplements
equations, we need to envision the numerous unknowns that can cause
equations to go awry; while remembering that a candle in the dark
is better than nothing at all. Pondering on the sources of uncertainty
and the tools we have to capture it, might lead us to believe that,
either, the level of our mathematical knowledge is not advanced enough,
or, we are using the wrong methods. The dichotomy between logic and
randomness is a topic for another time.
\end{doublespace}

\begin{doublespace}

\section{Methodological Fundamentals}
\end{doublespace}

\begin{doublespace}

\subsection{Notation and Terminology for Key Results}
\end{doublespace}
\begin{itemize}
\begin{doublespace}
\item $D_{BC}\left(p_{i},p_{i}^{\prime}\right)$, the Bhattacharyya Distance
between two multinomial populations each consisting of $k$ categories
classes with associated probabilities $p_{1},p_{2},...,p_{k}$ and
$p_{1}^{\prime},p_{2}^{\prime},...,p_{k}^{\prime}$ respectively.
\item $\rho\left(p_{i},p_{i}^{\prime}\right)$, the Bhattacharyya Coefficient.
\item $D_{BC-N}(p,q)$ is the Bhattacharyya distance between $p$ and $q$
normal distributions or classes.
\item $D_{BC-MN}\left(p_{1},p_{2}\right)$ is the Bhattacharyya distance
between two multivariate normal distributions, $\boldsymbol{p_{1}},\boldsymbol{p_{2}}$
where $\boldsymbol{p_{i}}\sim\mathcal{N}(\boldsymbol{\mu}_{i},\,\boldsymbol{\Sigma}_{i})$.
\item $D_{BC-TN}(p,q)$ is the Bhattacharyya distance between $p$ and $q$
truncated normal distributions or classes.
\item $D_{BC-TMN}\left(p_{1},p_{2}\right)$ is the Bhattacharyya distance
between two truncated multivariate normal distributions, $\boldsymbol{p_{1}},\boldsymbol{p_{2}}$
where $\boldsymbol{p_{i}}\sim\mathcal{N}(\boldsymbol{\mu}_{i},\,\boldsymbol{\Sigma}_{i},\,\boldsymbol{a}_{i},\,\boldsymbol{b}_{i})$.\end{doublespace}

\end{itemize}
\begin{doublespace}

\subsection{Bhattacharyya Distance}
\end{doublespace}

\begin{doublespace}
We use the Bhattacharyya distance (Bhattacharyya 1943, 1946) as a
measure of similarity or dissimilarity between the probability distributions
of the two entities we are looking to compare. These entities could
be two securities, groups of securities, markets or any statistical
populations that we are interested in studying. The Bhattacharyya
distance is defined as the negative logarithm of the Bhattacharyya
coefficient. 
\[
D_{BC}\left(p_{i},p_{i}^{\prime}\right)=-\ln\left[\rho\left(p_{i},p_{i}^{\prime}\right)\right]
\]
The Bhattacharyya coefficient is calculated as shown below for discrete
and continuous probability distributions. 
\[
\rho\left(p_{i},p_{i}^{\prime}\right)=\sum_{i}^{k}\sqrt{p_{i}p_{i}^{\prime}}
\]
\[
\rho\left(p_{i},p_{i}^{\prime}\right)=\int\sqrt{p_{i}\left(x\right)p_{i}^{\prime}\left(x\right)}dx
\]

Bhattacharyya\textquoteright s original interpretation of the measure
was geometric (Derpanis 2008). He considered two multinomial populations
each consisting of $k$ categories classes with associated probabilities
$p_{1},p_{2},...,p_{k}$ and $p_{1}^{\prime},p_{2}^{\prime},...,p_{k}^{\prime}$
respectively. Then, as $\sum_{i}^{k}p_{i}=1$ and $\sum_{i}^{k}p_{i}^{\prime}=1$,
he noted that $(\sqrt{p_{1}},...,\sqrt{p_{k}})$ and $(\sqrt{p_{1}^{\prime}},...,\sqrt{p_{k}^{\prime}})$
could be considered as the direction cosines of two vectors in $k-$dimensional
space referred to a system of orthogonal co-ordinate axes. As a measure
of divergence between the two populations Bhattacharyya used the square
of the angle between the two position vectors. If $\theta$ is the
angle between the vectors then: 
\[
\rho\left(p_{i},p_{i}^{\prime}\right)=cos\theta=\sum_{i}^{k}\sqrt{p_{i}p_{i}^{\prime}}
\]
Thus if the two populations are identical: $cos\theta=1$ corresponding
to $\theta=0$, hence we see the intuitive motivation behind the definition
as the vectors are co-linear. Bhattacharyya further showed that by
passing to the limiting case a measure of divergence could be obtained
between two populations defined in any way given that the two populations
have the same number of variates. The value of coefficient then lies
between $0$ and $1$. 
\[
0\leq\rho\left(p_{i},p_{i}^{\prime}\right)\leq=1
\]
\[
0\leq D_{BC}\left(p_{i},p_{i}^{\prime}\right)\leq\infty
\]
We get the following formulae (Lee and Bretschneider 2012) for the
Bhattacharyya distance when applied to the case of two uni-variate
normal distributions. 
\[
D_{BC-N}(p,q)=\frac{1}{4}\ln\left(\frac{1}{4}\left(\frac{\sigma_{p}^{2}}{\sigma_{q}^{2}}+\frac{\sigma_{q}^{2}}{\sigma_{p}^{2}}+2\right)\right)+\frac{1}{4}\left(\frac{(\mu_{p}-\mu_{q})^{2}}{\sigma_{p}^{2}+\sigma_{q}^{2}}\right)
\]

$\sigma_{p}$ is the variance of the $p-$th distribution, 

$\mu_{p}$ is the mean of the $p-$th distribution, and 

$p,q$ are two different distributions.

The original paper on the Bhattacharyya distance (Bhattacharyya 1943)
mentions a natural extension to the case of more than two populations.
For an $M$ population system, each with $k$ random variates, the
definition of the coefficient becomes, 
\[
\rho\left(p_{1},p_{2},...,p_{M}\right)=\int\cdots\int\left[p_{1}\left(x\right)p_{2}\left(x\right)...p_{M}\left(x\right)\right]^{\frac{1}{M}}dx_{1}\cdots dx_{k}
\]

For two multivariate normal distributions, $\boldsymbol{p_{1}},\boldsymbol{p_{2}}$
where $\boldsymbol{p_{i}}\sim\mathcal{N}(\boldsymbol{\mu}_{i},\,\boldsymbol{\Sigma}_{i}),$
\[
D_{BC-MN}\left(p_{1},p_{2}\right)=\frac{1}{8}(\boldsymbol{\mu}_{1}-\boldsymbol{\mu}_{2})^{T}\boldsymbol{\Sigma}^{-1}(\boldsymbol{\mu}_{1}-\boldsymbol{\mu}_{2})+\frac{1}{2}\ln\,\left(\frac{\det\boldsymbol{\Sigma}}{\sqrt{\det\boldsymbol{\Sigma}_{1}\,\det\boldsymbol{\Sigma}_{2}}}\right),
\]

$\boldsymbol{\mu}_{i}$ and $\boldsymbol{\Sigma}_{i}$ are the means
and covariances of the distributions, and $\boldsymbol{\Sigma}=\frac{\boldsymbol{\Sigma}_{1}+\boldsymbol{\Sigma}_{2}}{2}$.
We need to keep in mind that a discrete sample could be stored in
matrices of the form $A$ and $B$, where, $n$ is the number of observations
and $m$ denotes the number of variables captured by the two matrices.
\[
\boldsymbol{A_{m\times n}}\sim\mathcal{N}\left(\boldsymbol{\mu_{1}},\boldsymbol{\varSigma_{1}}\right)
\]
\[
\boldsymbol{B_{m\times n}}\sim\mathcal{N}\left(\boldsymbol{\mu_{2}},\boldsymbol{\varSigma_{2}}\right)
\]

\end{doublespace}

\begin{doublespace}

\subsection{Dimension Reduction}
\end{doublespace}

\begin{doublespace}
A key requirement to apply the Bhattacharyya distance in practice
is to have data-sets with the same number of dimensions. (Fodor 2002;
Burges 2009; Sorzano, Vargas and Montano 2014) are comprehensive collections
of methodologies aimed at reducing the dimensions of a data-set using
Principal Component Analysis or Singular Value Decomposition and related
techniques. (Johnson and Lindenstrauss 1984) proved a fundamental
result (JL Lemma) that says that any $n$ point subset of Euclidean
space can be embedded in $k=O(log\frac{n}{\epsilon^{2}})$ dimensions
without distorting the distances between any pair of points by more
than a factor of $\left(1\pm\epsilon\right)$, for any $0<\epsilon<1$.
Whereas principal component analysis is only useful when the original
data points are inherently low dimensional, the JL Lemma requires
absolutely no assumption on the original data. Also, note that the
final data points have no dependence on $d$, the dimensions of the
original data which could live in an arbitrarily high dimension. We
use the version of the bounds for the dimensions of the transformed
subspace given in (Frankl and Maehara 1988; 1990; Dasgupta and Gupta
1999).
\end{doublespace}
\begin{lem}
\begin{doublespace}
\label{Prop:Johnson and Lindenstrauss --- Dasgupta and Gupta}For
any $0<\epsilon<1$ and any integer $n$, let $k$ be a positive integer
such that 
\[
k\geq4\left(\frac{\epsilon^{2}}{2}-\frac{\epsilon^{3}}{3}\right)^{-1}\ln n
\]
Then for any set $V$ of $n$ points in $\boldsymbol{R}^{d}$, there
is a map $f:\boldsymbol{R}^{d}\rightarrow\boldsymbol{R}^{k}$ such
that for all $u,v\in V$, 
\[
\left(1-\epsilon\right)\Vert u-v\Vert^{2}\leq\Vert f\left(u\right)-f\left(v\right)\Vert^{2}\leq\left(1+\epsilon\right)\Vert u-v\Vert^{2}
\]
Furthermore, this map can be found in randomized polynomial time and
one such map is $f=\frac{1}{\sqrt{k}}Ax$ where, $x\in\boldsymbol{R}^{d}$
and $A$ is a $k\times d$ matrix in which each entry is sampled i.i.d
from a Gaussian $N\left(0,1\right)$ distribution.\end{doublespace}

\end{lem}
\begin{doublespace}
(Kashyap 2016b) provides expressions for the density functions after
dimension transformation when considering log normal distributions,
truncated normal and truncated multivariate normal distributions (Appendix
A: \ref{sec:Appendix-A:-Dimension Reduction, Distance Measures and Covariance}).
These results are applicable in the context of many variables observed
in real life such as stock prices, heart rates and volatilities, which
do not take on negative values. For completeness, we also include
the expression for the dimension transformed normal distribution.
A relationship between covariance and distance measures is also derived.
An asset pricing and one biological application show the limitless
possibilities such a comparison affords. Some pointers for implementation
and R code snippets for the Johnson Lindenstrauss matrix transformation
and a modification to the routine currently available to calculate
the Bhattacharyya distance are also listed. This modification allows
much larger numbers and dimensions to be handled, by utilizing the
properties of logarithms and the eigen values of a matrix.
\end{doublespace}

\begin{doublespace}

\section{From Symbols to Numbers, Empirical Illustrations across Markets}
\end{doublespace}

\begin{doublespace}
We illustrate several examples of how this measure could be used to
compare different countries based on the time series variables across
all equity securities traded in that market. Our data sample contains
prices (open, close, high and low) and trading volumes for most of
the securities from six different markets from Jan 01, 2014 to May
28, 2014 (Figure \ref{fig:Markets-and-Tickers}). Singapore with 566
securities is the market with the least number of traded securities.
Even if we reduce the dimension of all the other markets with more
number of securities, for a proper comparison of these markets, we
would need more than two years worth of data. Hence as a simplification,
we first reduce the dimension of the matrix holding the prices or
volumes for each market using principal component analysis (PCA; see
Shlens 2014) reduction, so that the number of tickers retained would
be comparable to the number of days for which we have data. We report
the results of using distance measures over the full sample after
PCA reduction.

We report the full matrix and not just the upper or lower matrix since
the PCA reduction we do takes the first country, reduces the dimensions
upto a certain number of significant digits and then reduces the dimension
of the second country to match the number of dimensions of the first
country. For example, this would mean that comparing AUS and SGP is
not exactly the same as comparing SGP and AUS. As a safety step before
calculating the distance, which requires the same dimensions for the
structures holding data for the two entities being compared, we could
perform dimension reduction using JL Lemma if the dimensions of the
two countries differs after the PCA reduction. We repeat the calculations
for different number of significant digits of the PCA reduction. This
shows the fine granularity of the results that our distance comparison
produces and highlights the issue that with PCA reduction there is
loss of information, since with different number of significant digits
employed in the PCA reduction, we get the result that different markets
are similar.

We illustrate another example, where we compare a randomly selected
sub universe of securities in each market, so that the number of tickers
retained would be comparable to the number of days for which we have
data. This approach could also be used when groups of securities are
being compared within the same market, a very common scenario when
deciding on the group of securities to invest in a market as opposed
to deciding which markets to invest in. Such an approach would be
highly useful for index construction or comparison across sectors
within a market.

We report the full matrix for the same reason as explained earlier
and perform multiple iterations when reducing the dimension using
the JL Lemma. A key observation is that the magnitude of the distances
are very different when using PCA reduction and when using dimension
reduction, due to the loss of information that comes with the PCA
technique. It is apparent that using dimension reduction via the JL
Lemma produces consistent results, since the same pairs of markets
are seen to be similar in different iterations. It is worth remembering
that in each iteration of the JL Lemma dimension transformation we
multiply by a different random matrix and hence the distance is slightly
different in each iteration but within the bound established by JL
Lemma. When the distance is somewhat close between two pairs of entities,
we could observe an inconsistency due to the JL Lemma transformation
in successive iterations. 

Lastly, we calculate sixty day moving volatilities on the close price
and trading volume and calculate the distance measure over the full
sample and also across each of the randomly selected sub-samples.
\end{doublespace}

\begin{doublespace}

\subsection{Speaking Volumes Of: Comparison of Trading Volumes}
\end{doublespace}

\begin{figure}[H]
\subfloat[Markets and Tickers Count\label{fig:Markets-and-Tickers}]{\includegraphics{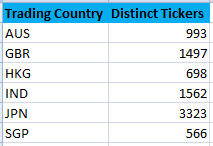}

}\hfill{}\subfloat[Volume PCA Dimension Reduction\label{fig:Volume-PCA-Results}]{\includegraphics[width=7cm,height=10cm]{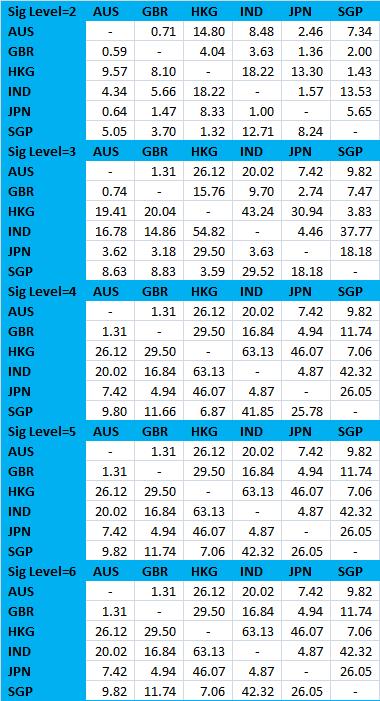}

}

\caption{Security Count by Market / Volume Distance Measures over Full Sample\label{fig:Security-Count-Volume-Distance}}

\end{figure}

\begin{doublespace}
The results of the volume comparison over the full sample are shown
in Figure \ref{fig:Volume-PCA-Results}. For example, in Figure \ref{fig:Volume-PCA-Results},
AUS - GBR are the most similar markets when two significant digits
are used and AUS - GBR are the most similar with six significant digits.
In this case the PCA and JL Lemma dimension reduction give similar
results.

\begin{figure}[H]
\subfloat[Volume PCA Dimension Reduction\label{fig:Volume-PCA-Dimension-Reduction}]{\includegraphics[width=7cm,height=10cm]{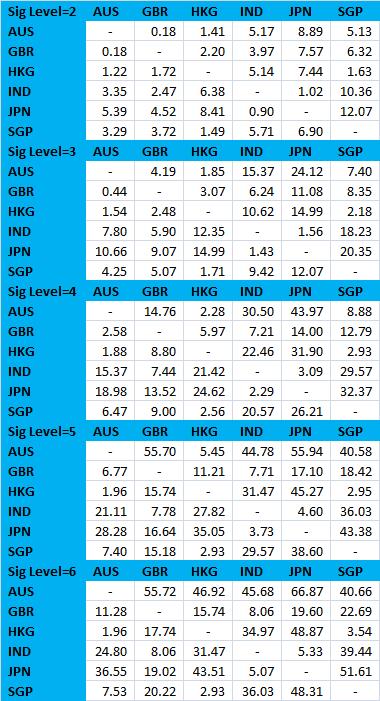}

}\hfill{}\subfloat[Volume JL Lemma Dimension Reduction\label{fig:Volume-JL-Lemma-Dimenion}]{\includegraphics[width=8cm,height=10cm]{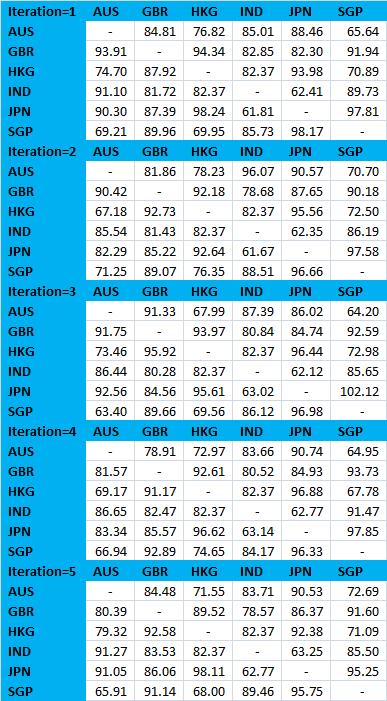}

}\caption{Volume Distance Measures over Randomly Chosen Sub Universe \label{fig:Volume-Results-with-Randomly}}

\end{figure}

The random sample results are shown in Figure \ref{fig:Volume-Results-with-Randomly}.
The left table (Figure \ref{fig:Volume-PCA-Dimension-Reduction})
is for PCA reduction on a randomly chosen sub universe and the right
table (Figure \ref{fig:Volume-JL-Lemma-Dimenion}) is for dimension
reduction using JL Lemma for the same sub universe.


\end{doublespace}

\begin{doublespace}

\subsection{A Pricey Prescription: Comparison of Prices (Open, Close, High and
Low)}
\end{doublespace}

\begin{doublespace}

\subsubsection{Open Close}
\end{doublespace}

\begin{figure}[H]
\subfloat[Open PCA Dimension Reduction\label{fig:Open-PCA-Dimension-Reduction}]{\includegraphics[width=7.5cm,height=10cm]{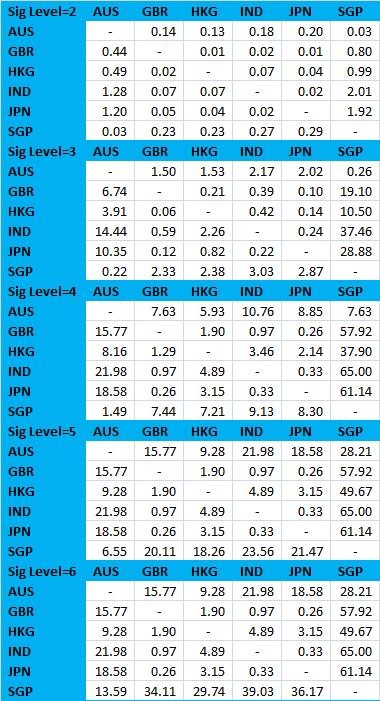}

}\hfill{}\subfloat[Close PCA Dimension Reduction\label{fig:Close-PCA-Dimension-Reduction}]{\includegraphics[width=7.5cm,height=10cm]{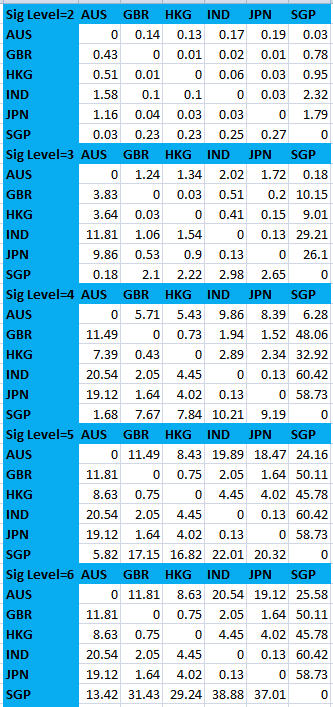}

}\caption{Open / Close Distance Measures over Full Sample\label{fig:Open-Close-Distance-Measures-Full Sample}}
\end{figure}

\begin{doublespace}
The results of a comparison between open and close prices over the
full sample are shown in Figures \ref{fig:Open-Close-Distance-Measures-Full Sample},
\ref{fig:Open-PCA-Dimension-Reduction}, \ref{fig:Close-PCA-Dimension-Reduction}.
For example, in Figure \ref{fig:Close-PCA-Dimension-Reduction}, AUS
- SGP are the most similar markets when two significant digits are
used and AUS - HKG are the most similar with six significant digits.
The similarities between open and close prices, in terms of the distance
measure, are also easily observed.

\begin{figure}[H]
\subfloat[Open PCA Dimension Reduction\label{fig:Open-PCA-Dimension-Reduction-Random}]{\includegraphics[width=7cm,height=10cm]{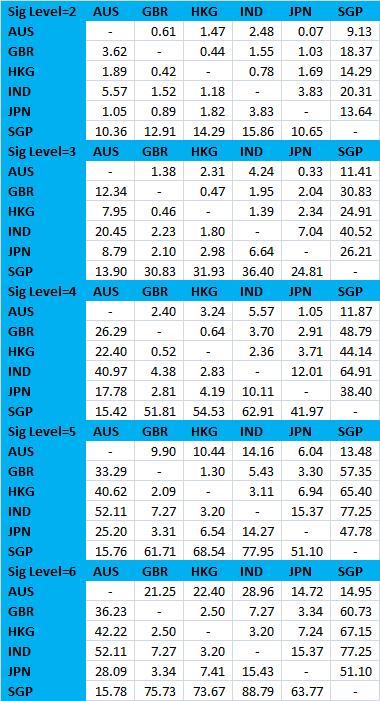}

}\hfill{}\subfloat[Open JL Lemma Dimension Reduction\label{fig:Open-JL-Lemma-Dimenion}]{\includegraphics[width=8cm,height=10cm]{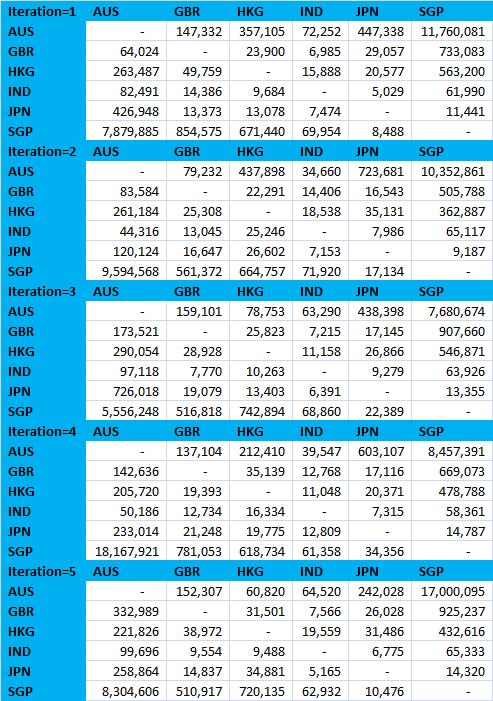}

}\caption{Open Distance Measures over Randomly Chosen Sub Universe \label{fig:Open-Results-with-Randomly}}
\end{figure}

\begin{figure}[H]
\subfloat[Close PCA Dimension Reduction\label{fig:Close-PCA-Dimension-Reduction-Random}]{\includegraphics[width=7cm,height=10cm]{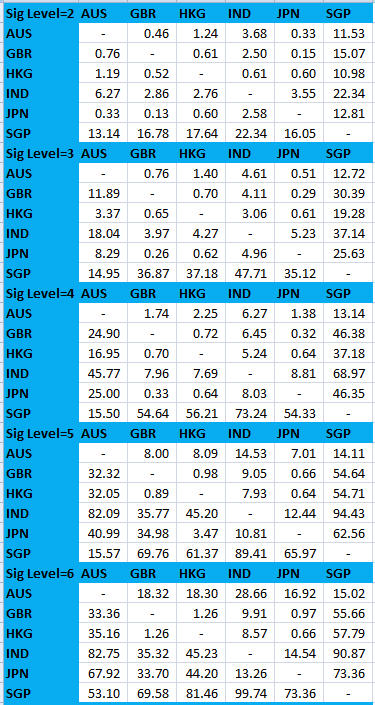}

}\hfill{}\subfloat[Close JL Lemma Dimension Reduction\label{fig:Close-JL-Lemma-Dimenion}]{\includegraphics[width=8cm,height=10cm]{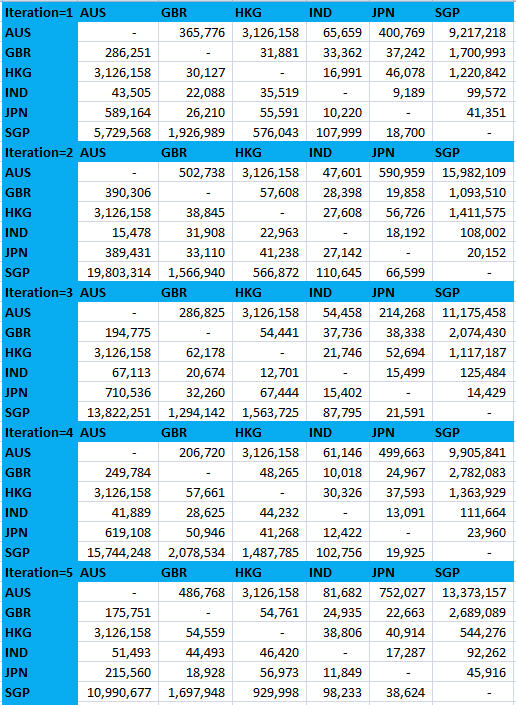}

}\caption{Close Distance Measures over Randomly Chosen Sub Universe \label{fig:Close-Results-with-Randomly}}
\end{figure}

The random sample results are shown in Figures \ref{fig:Open-Results-with-Randomly},
\ref{fig:Close-Results-with-Randomly}. The left table (Figures \ref{fig:Open-PCA-Dimension-Reduction-Random},
\ref{fig:Close-PCA-Dimension-Reduction-Random}) is for PCA reduction
on a randomly chosen sub universe and the right table (Figures \ref{fig:Open-JL-Lemma-Dimenion},
\ref{fig:Close-JL-Lemma-Dimenion}) is for dimension reduction using
JL Lemma for the same sub universe. In Figure \ref{fig:Close-JL-Lemma-Dimenion},
AUS - IND are the most similar in iteration one and also in iteration
five.

\end{doublespace}

\begin{doublespace}

\subsubsection{High Low}
\end{doublespace}

\begin{figure}[H]
\subfloat[High PCA Dimension Reduction\label{fig:High-PCA-Dimension-Reduction}]{\includegraphics[width=7.5cm,height=10cm]{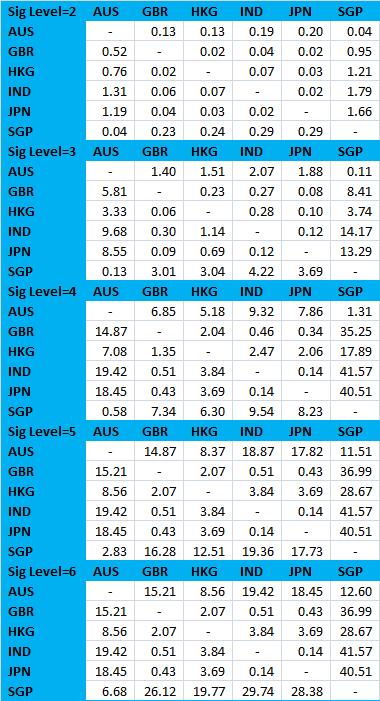}

}\hfill{}\subfloat[Low PCA Dimension Reduction\label{fig:Low-PCA-Dimension-Reduction}]{\includegraphics[width=7.5cm,height=10cm]{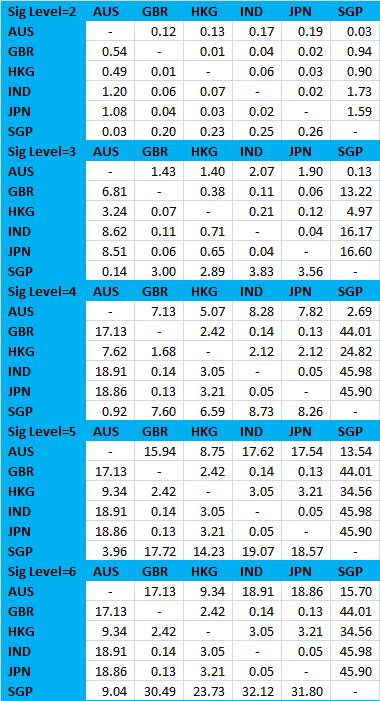}

}\caption{High / Low Distance Measures over Full Sample\label{fig:High-Low-Distance-Measures-Full Sample}}
\end{figure}

\begin{doublespace}
The results of a comparison between high and low prices over the full
sample are shown in Figures \ref{fig:High-Low-Distance-Measures-Full Sample},
\ref{fig:High-PCA-Dimension-Reduction}, \ref{fig:Low-PCA-Dimension-Reduction}.
For example, in Figure \ref{fig:High-PCA-Dimension-Reduction}, AUS
- SGP are the most similar markets when two significant digits are
used and AUS - HKG are the most similar with six significant digits.
The similarities between high and low prices are also easily observed.

\begin{figure}[H]
\subfloat[High PCA Dimension Reduction\label{fig:High-PCA-Dimension-Reduction-Random}]{\includegraphics[width=7cm,height=10cm]{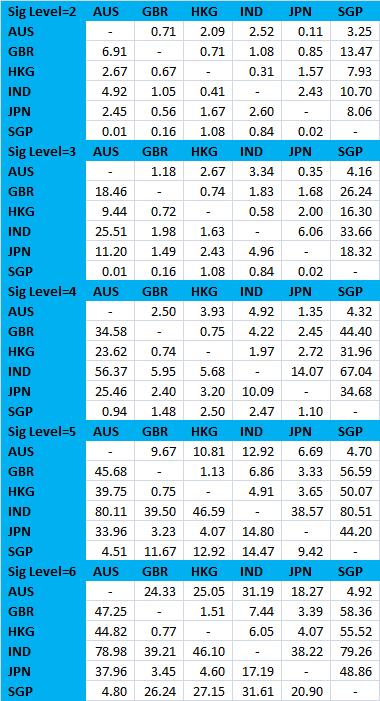}

}\hfill{}\subfloat[High JL Lemma Dimension Reduction\label{fig:High-JL-Lemma-Dimenion}]{\includegraphics[width=8cm,height=10cm]{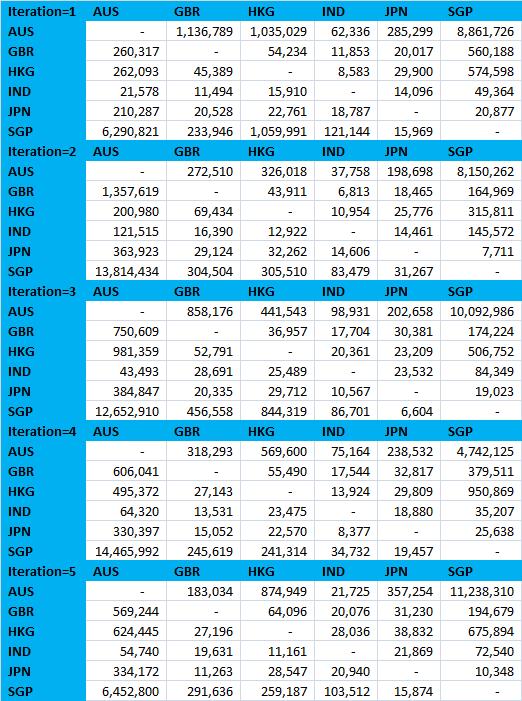}

}\caption{High Distance Measures over Randomly Chosen Sub Universe \label{fig:High-Results-with-Randomly}}
\end{figure}

\begin{figure}[H]
\subfloat[Low PCA Dimension Reduction\label{fig:Low-PCA-Dimension-Reduction-Random}]{\includegraphics[width=7cm,height=10cm]{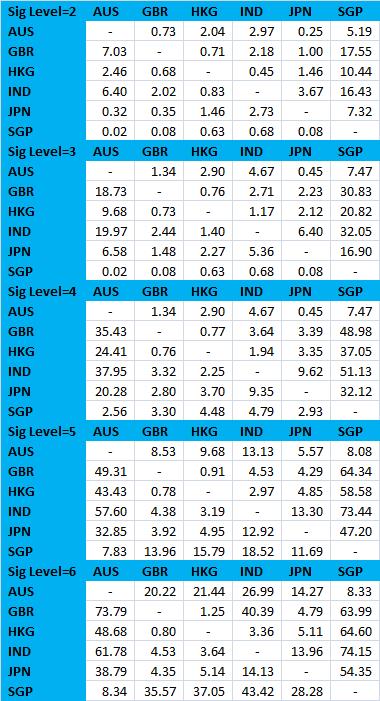}

}\hfill{}\subfloat[Low JL Lemma Dimension Reduction\label{fig:Low-JL-Lemma-Dimension}]{\includegraphics[width=8cm,height=10cm]{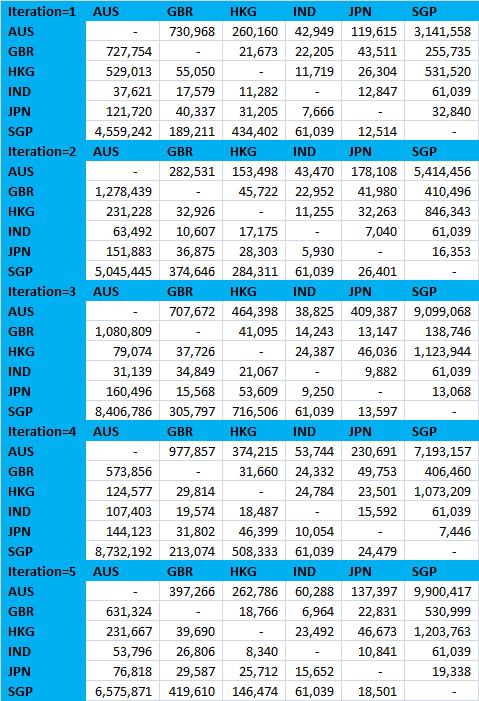}

}\caption{Low Distance Measures over Randomly Chosen Sub Universe \label{fig:Low-Results-with-Randomly}}
\end{figure}

The random sample results are shown in Figures \ref{fig:High-Results-with-Randomly},
\ref{fig:Low-Results-with-Randomly}. The left table (Figures \ref{fig:High-PCA-Dimension-Reduction-Random},
\ref{fig:Low-PCA-Dimension-Reduction-Random}) is for PCA reduction
on a randomly chosen sub universe and the right table (Figures \ref{fig:High-JL-Lemma-Dimenion},
\ref{fig:Low-JL-Lemma-Dimension}) is for dimension reduction using
JL Lemma for the same sub universe. In Figures \ref{fig:High-JL-Lemma-Dimenion}
and \ref{fig:Low-JL-Lemma-Dimension}, AUS - IND are the most similar
in iteration one and also in iteration five.

\end{doublespace}

\begin{doublespace}

\subsection{Taming the (Volatility) Skew: Comparison of Close Price / Volume
Volatilities}
\end{doublespace}

\begin{figure}[H]
\subfloat[Close Volatility PCA Dimension Reduction\label{fig:Close-Volatility-PCA-Dimension-Reduction}]{\includegraphics[width=7.5cm,height=10cm]{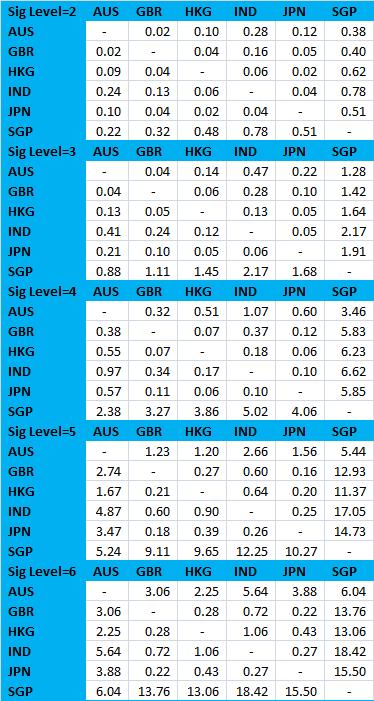}

}\hfill{}\subfloat[Volume Volatility PCA Dimension Reduction\label{fig:Volume-Volatility-PCA-Dimension-Reduction}]{\includegraphics[width=7.5cm,height=10cm]{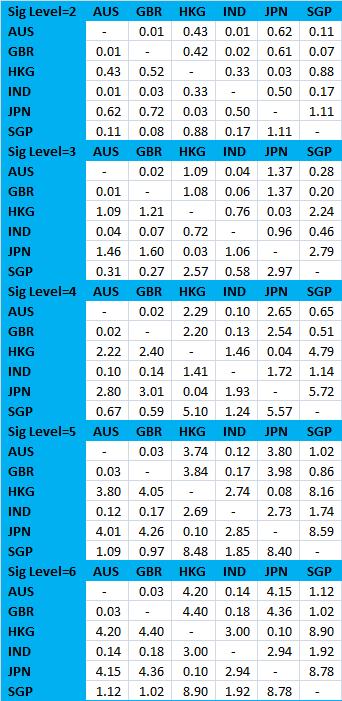}

}\caption{Close / Volume Volatility Distance Measures over Full Sample\label{fig:Close-Volume-Volatility-Distance-Measures-Full Sample}}
\end{figure}

\begin{doublespace}
The results of a comparison between close price volatilities and volume
volatilities over the full sample are shown in Figures \ref{fig:Close-Volume-Volatility-Distance-Measures-Full Sample},
\ref{fig:Close-Volatility-PCA-Dimension-Reduction}, \ref{fig:Volume-Volatility-PCA-Dimension-Reduction}.
For example, in Figure \ref{fig:Close-Volatility-PCA-Dimension-Reduction},
AUS - GBR are the most similar markets when two significant digits
are used and AUS - HKG are the most similar with six significant digits.
In Figure \ref{fig:Volume-Volatility-PCA-Dimension-Reduction}, AUS
- GBR - IND are equally similar markets when two significant digits
are used and AUS - GBR are the most similar with six significant digits.
The difference in magnitudes of the distance measures for prices,
volumes and volatilities is also easily observed. What this indicates
is that, prices are from the most dissimilar or distant distributions,
volatilities are less similar and volumes are from the most similar
or overlapping distributions. As also observed in the volume comparisons,
volume volatility comparisons give seemingly similar results when
PCA or JL Lemma dimension reductions are used. By considering the
price volatilities, and creating portfolios of instruments that have
dissimilar volatility distributions, we could reduce the overall risk
or variance of the portfolio returns, becoming one potential way of
mitigating the effects of wild volatility swings.

\begin{figure}[H]
\subfloat[Close Volatility PCA Dimension Reduction\label{fig:Close-Volatility-PCA-Dimension-Reduction-Random}]{\includegraphics[width=7cm,height=10cm]{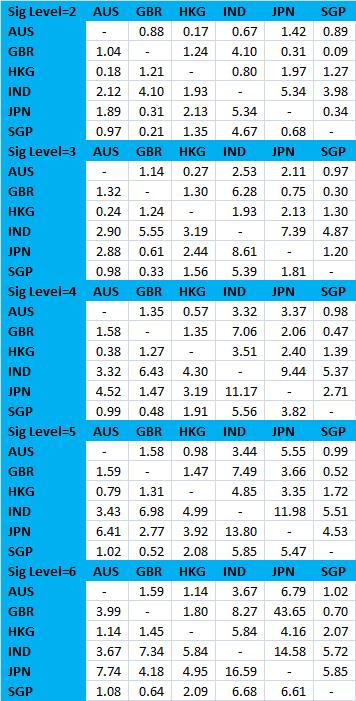}

}\hfill{}\subfloat[Close Volatility JL Lemma Dimension Reduction\label{fig:Close-Volatility-JL-Lemma-Dimension}]{\includegraphics[width=8cm,height=10cm]{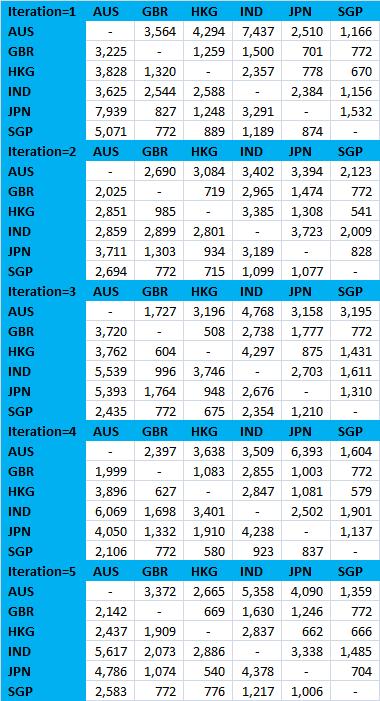}

}\caption{Close Volatility Distance Measures over Randomly Chosen Sub Universe
\label{fig:Close-Volatility-Results-with-Randomly}}
\end{figure}

\begin{figure}[H]
\subfloat[Volume Volatility PCA Dimension Reduction\label{fig:Volume-Volatility-PCA-Dimension-Reduction-Random}]{\includegraphics[width=7cm,height=10cm]{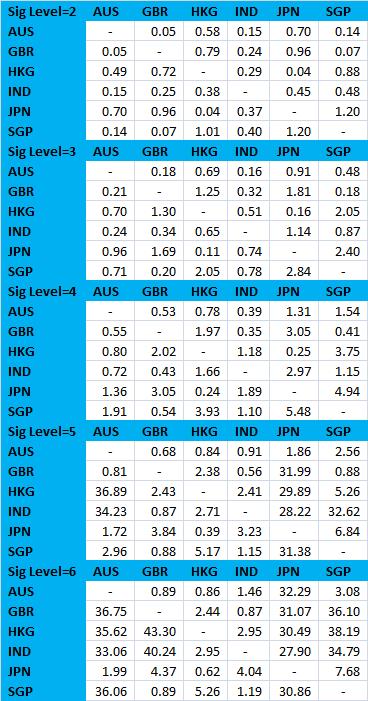}

}\hfill{}\subfloat[Volume Volatility JL Lemma Dimension Reduction\label{fig:Volume-Volatility-JL-Lemma-Dimension}]{\includegraphics[width=8cm,height=10cm]{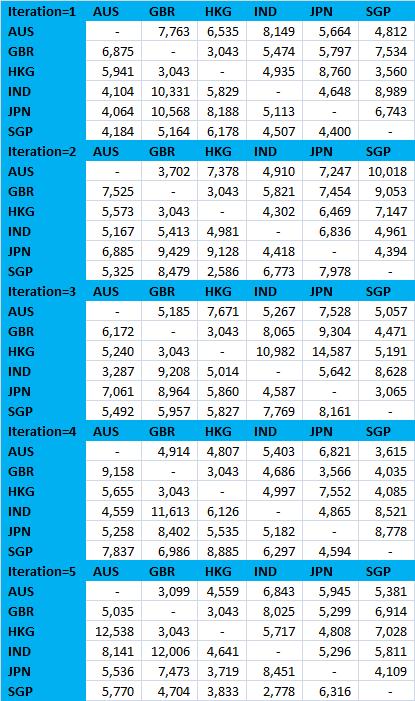}

}\caption{Volume Volatility Distance Measures over Randomly Chosen Sub Universe
\label{fig:Volume-Volatility-Results-with-Randomly}}
\end{figure}

The random sample results are shown in Figures \ref{fig:Close-Volatility-Results-with-Randomly},
\ref{fig:Volume-Volatility-Results-with-Randomly}. The left table
(Figures \ref{fig:Close-Volatility-PCA-Dimension-Reduction-Random},
\ref{fig:Volume-Volatility-PCA-Dimension-Reduction-Random}) is for
PCA reduction on a randomly chosen sub universe and the right table
(Figures \ref{fig:Close-Volatility-JL-Lemma-Dimension}, \ref{fig:Volume-Volatility-JL-Lemma-Dimension})
is for dimension reduction using JL Lemma for the same sub universe.
In Figure \ref{fig:Close-Volatility-JL-Lemma-Dimension} AUS - SGP
are the most similar in iteration one and also in iteration five.
In Figure \ref{fig:Volume-Volatility-JL-Lemma-Dimension}, AUS - SGP
are the most similar in iteration one and AUS- GBR in iteration five.

\end{doublespace}

\begin{doublespace}

\section{Possibilities for Future Research}
\end{doublespace}
\begin{enumerate}
\begin{doublespace}
\item A key limitation of this study is that we have reduced dimensions
using PCA or randomly sampled a sub matrix from the overall data-set
so that the length of time series available is in the range of the
number of securities that could be compared. Using a longer time series
for the variables would be a useful extension and a real application
would benefit immensely from more history.
\item We have used the simple formula for the Bhattacharyya distance applicable
to multivariate normal distributions. The formulae we have developed
over a truncated multivariate normal distribution or using a Normal
Log-Normal Mixture could give more accurate results. Again, later
works should look into tests that can establish which of the distributions
would apply depending on the data-set under consideration.
\item For each market we have looked at seven variables, open, close, low,
high, volume, close volatility and volume volatility. These variables
can be combined using the expression for the multinomial distance
to get a complete representation of which markets are more similar
than others. We aim to develop this methodology and illustrate these
techniques further in later works.
\item Once we have the similarity measures across groups of securities,
portfolios could be constructed to see how sensitive they are to different
explanatory factors and then performance benchmarks could be used
to guage the risk return relationship.\end{doublespace}

\end{enumerate}
\begin{doublespace}

\section{Conclusions }
\end{doublespace}

\begin{doublespace}
We have discussed how the combination of the Bhattacharyya distance
and the Johnson Lindenstrauss Lemma provides us with a practical and
novel methodology that allows comparisons between any two probability
distributions. This approach can help in the comparison of systems
that generate prices, quantities and aid in the analysis of shopping
patterns and understanding consumer behavior. The systems could be
goods transacted at different shopping malls or trade flows across
entire countries. Study of traffic congestion, road accidents and
other fatalities across two regions could be performed to get an idea
of similarities and seek common answers where such simplifications
might be applicable. Clearly, this methodology lends itself to numerous
applications outside the realm of finance and economics.

We have illustrated the comparison of prices, volumes and volatilities
across six different markets from three continents demonstrating the
power this methodology holds for big (small?) picture decision making. 

In Indian mythology (End-note \ref{enu:Avatar or Reincarnation};
Zimmer 1972; Doniger 1976; Rao 1993; Flood 1996; Parrinder 1997; Swami
2011), it is believed that in each era, God takes on an avatar or
reincarnation to fight the main source of evil in that epoch and to
restore the balance between good and bad. In this age of too much
information and complexity, perhaps the supreme being needs to be
born as a data scientist, conceivably with an apt superhero nickname,
the Infoman (For society's fascination with superheroes or superhumans
see: Eco and Chilton 1972; Reynolds 1992; Fingeroth 2004; Haslem,
Ndalianis and Mackie 2007; Coogan 2009). Until higher powers intervene
and provide the ultimate solution to completely eliminate information
overload, we have to make do with marginal methods to reduce information,
such as this composition.

As we wait for the perfect solution, it is worth meditating upon what
superior beings would do when faced with a complex situation, such
as the one we are in. It is said that the Universe is but the Brahma's
(Creator's) dream (Barnett 1907; Ramamurthi 1995; Ghatage 2010). Research
(Effort / Struggle) can help us understand this world; Sleep (Ease
/ Peace of Mind) can help us create our own world. We just need to
be mindful that the most rosy and well intentioned dreams can have
unintended consequences (Kashyap 2016e) and turn to nightmares (Nolan
2010; Lehrer 2010; Kashyap 2016f). 

Native to Australia (End-note \ref{enu:Down Under}), ``Koalas spend
about 4.7 hours eating, 4 minutes traveling, 4.8 hours resting while
awake and 14.5 hours sleeping in a 24-hour period'' - (Nagy and Martin1985).
See also (Smith 1979; Moyal 2008). The benefits of yoga on sleep quality
are well documented (End-note \ref{enu:Yoga}; Cohen, Warneke, Fouladi,
Rodriguez and Chaoul-Reich 2004; Khalsa 2004; Manjunath and Telles
2005; Chen, Chen, Chao, Hung, Lin and Li 2009; Vera, Manzaneque, Maldonado,
Carranque, Rodriguez, Blanca and Morell 2009).

\begin{figure}[H]
\includegraphics[width=17cm]{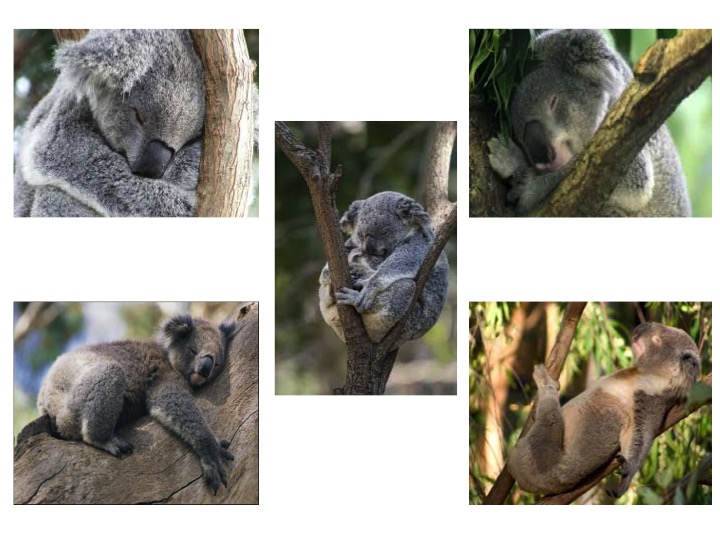}

\caption{Sleeping Like A Koala\label{fig:Sleeping-Like-A-Koala}}

\end{figure}

A lesson from close by and down under: We need to \textquotedblleft Do
Some Yoga and Sleep Like A Koala\textquotedblright{} (Figure \ref{fig:Sleeping-Like-A-Koala}).
With that, we present a list of sleeping aids in section \ref{sec:Sleeping-Aids}.

\end{doublespace}

\begin{doublespace}

\section{\label{sec:Sleeping-Aids}Sleeping Aids (Notes and References)}
\end{doublespace}
\begin{enumerate}
\begin{doublespace}
\item Dr. Yong Wang, Dr. Isabel Yan, Dr. Vikas Kakkar, Dr. Fred Kwan, Dr.
William Case, Dr. Srikant Marakani, Dr. Qiang Zhang, Dr. Costel Andonie,
Dr. Jeff Hong, Dr. Guangwu Liu, Dr. Humphrey Tung and Dr. Xu Han at
the City University of Hong Kong provided advice and more importantly
encouragement to explore and where possible apply cross disciplinary
techniques. The views and opinions expressed in this article, along
with any mistakes, are mine alone and do not necessarily reflect the
official policy or position of either of my affiliations or any other
agency.
\item \label{enu:The-Red-Queen's}The Red Queen's race is an incident that
appears in Lewis Carroll's Through the Looking-Glass and involves
the Red Queen, a representation of a Queen in chess, and Alice constantly
running but remaining in the same spot.
\end{doublespace}

\begin{doublespace}
\textquotedbl{}Well, in our country,\textquotedbl{} said Alice, still
panting a little, \textquotedbl{}you'd generally get to somewhere
else, if you run very fast for a long time, as we've been doing.\textquotedbl{}

\textquotedbl{}A slow sort of country!\textquotedbl{} said the Queen.
\textquotedbl{}Now, here, you see, it takes all the running you can
do, to keep in the same place. If you want to get somewhere else,
you must run at least twice as fast as that!\textquotedbl{}

\href{https://en.wikipedia.org/wiki/Red_Queen\%27s_race}{The Red Queen's Race, Wikipedia Link}
\end{doublespace}

This quote is commonly attributed as being from Alice in Wonderland
as: \textquotedblleft My dear, here we must run as fast as we can,
just to stay in place. And if you wish to go anywhere you must run
twice as fast as that.\textquotedblright{}

\begin{doublespace}
\item \label{enu:Portfolio Manager}\href{https://en.wikipedia.org/wiki/Portfolio_manager}{Portfolio Manager, Wikipedia Link}
\item \label{enu:Universal Computing Machine}\href{https://en.wikipedia.org/wiki/Universal_Turing_machine}{Universal Computing Machine, Wikipedia Link}
\item \label{enu:Computer}\href{https://en.wikipedia.org/wiki/Computer}{Computer, Wikipedia Link}
\item \label{enu: Mac or Macintosh}\href{https://en.wikipedia.org/wiki/Macintosh}{MAC or Macintosh, Wikipedia Link}
\item \label{enu:Personal Computer}\href{https://en.wikipedia.org/wiki/Personal_computer}{Personal Computer, Wikipedia Link}
\item \label{enu:MAC vs MPC}\href{https://en.wikipedia.org/wiki/Apple_Computer,_Inc._v._Microsoft_Corp.}{MAC vs MPC, Wikipedia Link}
\item \label{enu:History Computing}\href{https://en.wikipedia.org/wiki/History_of_computing}{History Computing, Wikipedia Link}
\item \label{enu:Computing Platform}\href{https://en.wikipedia.org/wiki/Computing_platform}{Computing Platform, Wikipedia Link}
\item \label{enu:Cloud Computing}\href{https://en.wikipedia.org/wiki/Cloud_computing}{Cloud Computing, Wikipedia Link}
\item \label{enu:Quantum Computing}\href{https://en.wikipedia.org/wiki/Quantum_computing}{Quantum Computing, Wikipedia Link}
\item \label{enu:Avatar or Reincarnation}\href{https://en.wikipedia.org/wiki/Avatar}{Avatar or Reincarnation, Wikipedia Link}
\item \label{enu:Yoga}\href{https://en.wikipedia.org/wiki/Yoga}{Yoga, Wikipedia Link}
\item \label{enu:Down Under}\href{https://en.wikipedia.org/wiki/Down_Under}{Australia or Down Under, Wikipedia Link}
\item Barnett, L. D. (1907). The Brahma Knowledge. An Outline of the Philosophy
of the Vedanta as Set Forth by the Upanishads and by Sankara. Wisdom
of the East series. E.P. Dutton Publishing, Boston, Massachusetts.
\item Bhattacharyya, A. (1943). On a Measure of Divergence Between Two Statistical
Populations Defined by their Probability Distributions, Bull. Calcutta
Math. Soc., 35, pp. 99-110.
\item Bhattacharyya, A. (1946). On a measure of divergence between two multinomial
populations. Sankhy\={a}: The Indian Journal of Statistics, 401-406.
\item Burges, C. J. (2009). Dimension reduction: A guided tour. Machine
Learning, 2(4), 275-365.
\item Burkardt, J. (2014). The Truncated Normal Distribution. Department
of Scientific Computing Website, Florida State University.
\item Carroll, L. (1865). (2012 Reprint) Alice's adventures in wonderland.
Random House, Penguin Random House, Manhattan, New York.
\item Carroll, L. (1871). (2009 Reprint) Through the looking glass: And
what Alice found there. Random House, Penguin Random House, Manhattan,
New York. 
\item Ceruzzi, P. E. (2003). A history of modern computing. MIT press.
\item Chen, K. M., Chen, M. H., Chao, H. C., Hung, H. M., Lin, H. S., \&
Li, C. H. (2009). Sleep quality, depression state, and health status
of older adults after silver yoga exercises: cluster randomized trial.
International journal of nursing studies, 46(2), 154-163.
\item Chiani, M., Dardari, D., \& Simon, M. K. (2003). New exponential bounds
and approximations for the computation of error probability in fading
channels. Wireless Communications, IEEE Transactions on, 2(4), 840-845.
\item Clark, P. K. (1973). A subordinated stochastic process model with
finite variance for speculative prices. Econometrica: journal of the
Econometric Society, 135-155.
\item Cody, W. J. (1969). Rational Chebyshev approximations for the error
function. Mathematics of Computation, 23(107), 631-637.
\item Cohen, L., Warneke, C., Fouladi, R. T., Rodriguez, M., \& Chaoul-Reich,
A. (2004). Psychological adjustment and sleep quality in a randomized
trial of the effects of a Tibetan yoga intervention in patients with
lymphoma. Cancer, 100(10), 2253-2260.\end{doublespace}

\item Coogan, P. (2009). The Definition of the Superhero. A comics studies
reader, 77.
\begin{doublespace}
\item Dasgupta, S., \& Gupta, A. (1999). An elementary proof of the Johnson-Lindenstrauss
lemma. International Computer Science Institute, Technical Report,
99-006.
\item Derpanis, K. G. (2008). The Bhattacharyya Measure. Mendeley Computer,
1(4), 1990-1992.
\item Doniger, W. (1976). The origins of evil in Hindu mythology (No. 6).
Univ of California Press.\end{doublespace}

\item Eco, U., \& Chilton, N. (1972). The myth of Superman.
\item Fingeroth, D. (2004). Superman on the Couch: What Superheroes Really
Tell Us about Ourselves and Our Society. A\&C Black.
\begin{doublespace}
\item Frankl, P., \& Maehara, H. (1988). The Johnson-Lindenstrauss lemma
and the sphericity of some graphs. Journal of Combinatorial Theory,
Series B, 44(3), 355-362.
\item Frankl, P., \& Maehara, H. (1990). Some geometric applications of
the beta distribution. Annals of the Institute of Statistical Mathematics,
42(3), 463-474.
\item Flood, G. D. (1996). An introduction to Hinduism. Cambridge University
Press.
\item Fodor, I. K. (2002). A survey of dimension reduction techniques. Technical
Report UCRL-ID-148494, Lawrence Livermore National Laboratory.
\item Ghatage, S. (2010). Brahma's Dream. Anchor Canada, Penguin Random
House, Manhattan, New York.\end{doublespace}

\item Haslem, W., Ndalianis, A., \& Mackie, C. J. (Eds.). (2007). Super/Heroes:
From Hercules to Superman. New Academia Publishing, LLC.
\begin{doublespace}
\item Horrace, W. C. (2005). Some results on the multivariate truncated
normal distribution. Journal of Multivariate Analysis, 94(1), 209-221.
\item Ifrah, G., Harding, E. F., Bellos, D., \& Wood, S. (2000). The universal
history of computing: From the abacus to quantum computing. John Wiley
\& Sons, Inc.\end{doublespace}

\item Johnson, W. B., \& Lindenstrauss, J. (1984). Extensions of Lipschitz
mappings into a Hilbert space. Contemporary mathematics, 26(189-206),
1.
\begin{doublespace}
\item Kashyap, R. (2014a). Dynamic Multi-Factor Bid\textendash Offer Adjustment
Model. The Journal of Trading, 9(3), 42-55.
\item Kashyap, R. (2014b). The Circle of Investment. International Journal
of Economics and Finance, 6(5), 244-263.
\item Kashyap, R. (2015a). Financial Services, Economic Growth and Well-Being:
A Four Pronged Study. Indian Journal of Finance, 9(1), 9-22.
\item Kashyap, R. (2015b). A Tale of Two Consequences. The Journal of Trading,
10(4), 51-95.
\item Kashyap, R. (2016a). Hong Kong - Shanghai Connect / Hong Kong - Beijing
Disconnect (?), Scaling the Great Wall of Chinese Securities Trading
Costs. The Journal of Trading, 11(3), 81-134.
\item Kashyap, R. (2016b). Combining Dimension Reduction, Distance Measures
and Covariance. Working Paper.
\item Kashyap, R. (2016c). Solving the Equity Risk Premium Puzzle and Inching
Towards a Theory of Everything. Working Paper.
\item Kashyap, R. (2016d). Fighting Uncertainty with Uncertainty. Working
Paper.
\item Kashyap, R. (2016e). Notes on Uncertainty, Unintended Consequences
and Everything Else. Working Paper.
\item Kashyap, R. (2016f). The American Dream, An Unsustainable Nightmare.
Working Paper.
\item Kattumannil, S. K. (2009). On Stein\textquoteright s identity and
its applications. Statistics \& Probability Letters, 79(12), 1444-1449.
\item Keynes, J. M. (1937). The General Theory of Employment. The Quarterly
Journal of Economics, 51(2), 209-223.
\item Keynes, J. M. (1971). The Collected Writings of John Maynard Keynes:
In 2 Volumes. A Treatise on Money. The Applied Theory of Money. Macmillan
for the Royal Economic Society. 
\item Keynes, J. M. (1973). A treatise on probability, the collected writings
of John Maynard Keynes, vol. VIII.
\item Khalsa, S. B. S. (2004). Treatment of chronic insomnia with yoga:
A preliminary study with sleep\textendash wake diaries. Applied psychophysiology
and biofeedback, 29(4), 269-278.
\item Kiani, M., Panaretos, J., Psarakis, S., \& Saleem, M. (2008). Approximations
to the normal distribution function and an extended table for the
mean range of the normal variables.
\item Kimeldorf, G., \& Sampson, A. (1973). A class of covariance inequalities.
Journal of the American Statistical Association, 68(341), 228-230.
\item Lawson, T. (1985). Uncertainty and economic analysis. The Economic
Journal, 95(380), 909-927.
\item Lee, K. Y., \& Bretschneider, T. R. (2012). Separability Measures
of Target Classes for Polarimetric Synthetic Aperture Radar Imagery.
Asian Journal of Geoinformatics, 12(2).
\item Lehrer, J. (2010). \href{https://www.wired.com/2010/07/the-neuroscience-of-inception/}{The Neuroscience of Inception}.
Wired 26 Jul. 2010. Web. 13 Aug. 2013.
\item Manjunath, N. K., \& Telles, S. (2005). Influence of Yoga \& Ayurveda
on self-rated sleep in a geriatric population. Indian Journal of Medical
Research, 121(5), 683.
\item Mazur, J. E. (2015). Learning and behavior. Psychology Press.
\item McManus, H., \& Hastings, D. (2005, July). 3.4. 1 A Framework for
Understanding Uncertainty and its Mitigation and Exploitation in Complex
Systems. In INCOSE International Symposium (Vol. 15, No. 1, pp. 484-503).
\item Mill, J. (1829). Analysis of the Phenomena of the Human Mind (Vol.
1, 2). Longmans, Green, Reader, and Dyer.
\item Miranda, M. J., \& Fackler, P. L. (2002). Applied Computational Economics
and Finance.
\item Moyal, A. (Ed.). (2008). Koala: a historical biography. CSIRO PUBLISHING.
\item Nagy, K. A., \& Martin, R. W. (1985). Field Metabolic Rate, Water
Flux, Food Consumption and Time Budget of Koalas, Phascolarctos Cinereus
(Marsupialia: Phascolarctidae) in Victoria. Australian Journal of
Zoology, 33(5), 655-665.
\item Nolan, C. (2010). Inception {[}film{]}. Warner Bros.: Los Angeles,
CA, USA.
\item Parrinder, E. G. (1997). Avatar and incarnation: the divine in human
form in the world's religions. Oneworld Publications Limited.
\item Ramamurthi, B. (1995). The fourth state of consciousness: The Thuriya
Avastha. Psychiatry and clinical neurosciences, 49(2), 107-110.
\item Rao, T. G. (1993). Elements of Hindu iconography. Motilal Banarsidass
Publisher.\end{doublespace}

\item Reynolds, R. (1992). Super heroes: A modern mythology. Univ. Press
of Mississippi.
\begin{doublespace}
\item Rubinstein, M. E. (1973). A comparative statics analysis of risk premiums.
The Journal of Business, 46(4), 605-615.
\item Rubinstein, M. (1976). The valuation of uncertain income streams and
the pricing of options. The Bell Journal of Economics, 407-425.\end{doublespace}

\item Shlens, J. (2014). A tutorial on principal component analysis. arXiv
preprint arXiv:1404.1100.
\begin{doublespace}
\item Simon, H. A. (1962). The Architecture of Complexity. Proceedings of
the American Philosophical Society, 106(6), 467-482.
\item Smith, M. (1979). Behaviour of the Koala, Phascolarctos Cinereus Goldfuss,
in Captivity. 1. Non-Social Behaviour. Wildlife Research, 6(2), 117-129.
\item Soranzo, A., \& Epure, E. (2014). Very simply explicitly invertible
approximations of normal cumulative and normal quantile function.
Applied Mathematical Sciences, 8(87), 4323-4341.
\item Sorzano, C. O. S., Vargas, J., \& Montano, A. P. (2014). A survey
of dimensionality reduction techniques. arXiv preprint arXiv:1403.2877.
\item Stein, C. M. (1973). Estimation of the mean of a multivariate normal
distribution. Proceedings of the Prague Symposium of Asymptotic Statistics.
\item Stein, C. M. (1981). Estimation of the mean of a multivariate normal
distribution. The annals of Statistics, 1135-1151.
\item Swami, B. (2011). Bhagavad Gita as it is. The Bhaktivedanta book trust,
Mumbai, India.
\item Tauchen, G. E., \& Pitts, M. (1983). The Price Variability-Volume
Relationship on Speculative Markets. Econometrica, 51(2), 485-505.
\item Teerapabolarn, K. (2013). Stein's identity for discrete distributions.
International Journal of Pure and Applied Mathematics, 83(4), 565.
\item Vera, F. M., Manzaneque, J. M., Maldonado, E. F., Carranque, G. A.,
Rodriguez, F. M., Blanca, M. J., \& Morell, M. (2009). Subjective
sleep quality and hormonal modulation in long-term yoga practitioners.
Biological psychology, 81(3), 164-168.
\item Williams, M. R. (1997). A history of computing technology. IEEE Computer
Society Press.
\item Yang, M. (2008). Normal log-normal mixture, leptokurtosis and skewness.
Applied Economics Letters, 15(9), 737-742.
\item Zimmer, H. R. (1972). Myths and symbols in Indian art and civilization
(Vol. 6). Princeton University Press.
\item Zogheib, B., \& Hlynka, M. (2009). Approximations of the Standard
Normal Distribution. University of Windsor, Department of Mathematics
and Statistics.\end{doublespace}

\end{enumerate}
\begin{doublespace}

\section{Appendix A: \label{sec:Appendix-A:-Dimension Reduction, Distance Measures and Covariance}Dimension
Reduction, Distance Measures and Covariance}
\end{doublespace}

\begin{doublespace}
All the results below are from (Kashyap 2016b). Other useful references
are pointed in the relevant sections below.
\end{doublespace}

\begin{doublespace}

\subsection{Normal Log-Normal Mixture}
\end{doublespace}

\begin{doublespace}
Transforming log-normal multi-variate variables into a lower dimension
by multiplication with an independent normal distribution (See Lemma
\ref{Prop:Johnson and Lindenstrauss --- Dasgupta and Gupta}) results
in the sum of variables with a normal log-normal mixture, (Clark 1973;
Tauchen and Pitts 1983; Yang 2008), evaluation of which requires numerical
techniques (Miranda and Fackler 2002).

A random variable, $U$, would be termed a normal log-normal mixture
if it is of the form,
\[
U=Xe^{Y}
\]
where, $X$ and $Y$ are random variables with correlation coefficient,
$\rho$ satisfying the below, 
\[
\left[\begin{array}{c}
X\\
Y
\end{array}\right]\sim N\left(\left[\begin{array}{c}
\mu_{X}\\
\mu_{Y}
\end{array}\right],\left[\begin{array}{cc}
\sigma_{X}^{2} & \rho\sigma_{X}\sigma_{Y}\\
\rho\sigma_{X}\sigma_{Y} & \sigma_{Y}^{2}
\end{array}\right]\right)
\]
We note that for $\sigma_{Y}=0$ when $Y$ degenerates to a constant,
this is just the distribution of $X$ and $\rho$ is unidentified.
\end{doublespace}

To transform a column vector with $d$ observations of a random variable
into a lower dimension of order, $k<d$, we can multiply the column
vector with a matrix, $A\sim N(0;\frac{1}{k})$ of dimension $k\times d$.
\begin{lem}
\begin{doublespace}
\label{prop:Normal_Lognormal_The-density-function} A dimension transformation
of $d$ observations of a log-normal variable into a lower dimension,
$k$, using Lemma \ref{Prop:Johnson and Lindenstrauss --- Dasgupta and Gupta},
yields a probability density function which is the sum of random variables
with a normal log-normal mixture, given by the convolution,
\[
f_{S}\left(s\right)=f_{U_{1}}\left(u_{1}\right)*f_{U_{2}}\left(u_{2}\right)*...*f_{U_{k}}\left(u_{k}\right)
\]
\[
\text{Here, }f_{U_{i}}\left(u_{i}\right)=\frac{\sqrt{k}}{2\pi\sigma_{Y_{i}}}\int_{-\infty}^{\infty}\;e^{-y-\frac{ku_{i}^{2}}{2e^{2y}}-\frac{\left[y-\mu_{Y_{i}}\right]^{2}}{2\sigma_{Y_{i}}^{2}}}dy
\]
\[
U_{i}=X_{i}e^{Y_{i}}
\]
\[
\left[\begin{array}{c}
X_{i}\\
Y_{i}
\end{array}\right]\sim N\left(\left[\begin{array}{c}
0\\
\mu_{Y_{i}}
\end{array}\right],\left[\begin{array}{cc}
\frac{1}{k} & 0\\
0 & \sigma_{Y_{i}}^{2}
\end{array}\right]\right)
\]
The convolution of two probability densities arises when we have the
sum of two independent random variables, $Z=X+Y$. The density of
$Z,\;h_{Z}\left(z\right)$ is given by,
\[
{\displaystyle h_{Z}\left(z\right)=\left(f_{X}\text{\textasteriskcentered}f_{Y}\right)\left(z\right)=f_{X}\left(x\right)*f_{Y}\left(y\right)=\int_{-\infty}^{\infty}f_{X}\left(z-y\right)*f_{Y}\left(y\right)dy=\int_{-\infty}^{\infty}f_{X}\left(x\right)*f_{Y}\left(z-x\right)dx}
\]
When the number of independent random variables being added is more
than two, or the reduced dimension after the Lemma \ref{Prop:Johnson and Lindenstrauss --- Dasgupta and Gupta}
transformation is more than two, $k>2$, then we can take the convolution
of the density resulting after the convolution of the first two random
variables, with the density of the third variable and so on in a pair
wise manner, till we have the final density.\end{doublespace}

\end{lem}

\subsection{Normal Normal Product}

For completeness, we illustrate how dimension reduction would work
on a data-set containing random variables that have normal distributions.
This can serve as a useful benchmark given the wide usage of the normal
distribution and can be an independently useful result, though most
variables observed in real life are normally not so normal.
\begin{lem}
\begin{doublespace}
\label{prop:Normal_Normal_The-density-function} A dimension transformation
of $d$ observations of a normal variable into a lower dimension,
$k$, using Lemma \ref{Prop:Johnson and Lindenstrauss --- Dasgupta and Gupta},
yields a probability density function which is the sum of random variables
with a normal normal product distribution, given by the convolution,
\[
f_{S}\left(s\right)=f_{U_{1}}\left(u_{1}\right)*f_{U_{2}}\left(u_{2}\right)*...*f_{U_{k}}\left(u_{k}\right)
\]
\[
\text{Here, }f_{U_{i}}\left(u_{i}\right)=\int_{-\infty}^{\infty}\left(\frac{1}{\left|x\right|}\right)\frac{1}{\sigma_{Y_{i}}\sqrt{2\pi}}\;e^{-\frac{\left(x-\mu_{Y_{i}}\right)^{2}}{2\sigma_{Y_{i}}^{2}}}\sqrt{\frac{k}{2\pi}}\;e^{-\frac{k\left(\frac{u_{i}}{x}\right)^{2}}{2}}dx
\]
\[
U_{i}=X_{i}Y_{i}
\]
\[
\left[\begin{array}{c}
X_{i}\\
Y_{i}
\end{array}\right]\sim N\left(\left[\begin{array}{c}
0\\
\mu_{Y_{i}}
\end{array}\right],\left[\begin{array}{cc}
\frac{1}{k} & 0\\
0 & \sigma_{Y_{i}}^{2}
\end{array}\right]\right)
\]
\end{doublespace}

\end{lem}
\begin{doublespace}

\subsection{Truncated Normal Distribution}
\end{doublespace}

\begin{doublespace}
A truncated normal distribution is the probability distribution of
a normally distributed random variable whose value is either bounded
below, above or both (Horrace 2005; Burkardt 2014). \textbf{(}Kiani,
Panaretos, Psarakis and Saleem 2008\textbf{; }Zogheib and Hlynka 2009;
Soranzo and Epure 2014) list some of the numerous techniques to calculate
the normal cumulative distribution. Approximations to the error function
are also feasible options (Cody 1969; Chiani, Dardari and Simon 2003).
Despite the truncation, this could be a potent extension when it is
known a-priori that the values a variable can take are almost surely
bounded.

Suppose $X\sim N(\mu,\sigma^{2})$ has a normal distribution and lies
within the interval $X\in(a,b),\;-\infty\leq a<b\leq\infty$. Then
$X$ conditional on $a<X<b$ has a truncated normal distribution.
Its probability density function, $f_{X}$, for $a\leq x\leq b$ ,
is given by
\[
f_{X}\left(x\mid\mu,\sigma^{2},a,b\right)=\begin{cases}
\frac{\frac{1}{\sigma}\phi\left(\frac{x-\mu}{\sigma}\right)}{\Phi\left(\frac{b-\mu}{\sigma}\right)-\Phi\left(\frac{a-\mu}{\sigma}\right)} & \quad;a\leq x\leq b\\
0 & ;\text{otherwise}
\end{cases}
\]

Here, ${\scriptstyle {\phi(\xi)=\frac{1}{\sqrt{2\pi}}\exp{(-\frac{1}{2}\xi^{2}})}}$
is the probability density function of the standard normal distribution
and ${\scriptstyle {\Phi(\cdot)}}$ is its cumulative distribution
function. There is an understanding that if ${\scriptstyle {b=\infty}\ }$,
then ${\scriptstyle {\Phi(\frac{b-\mu}{\sigma})=1}}$, and similarly,
if ${\scriptstyle {a=-\infty}}$ , then ${\scriptstyle {\Phi(\frac{a-\mu}{\sigma})=0}}$. 
\end{doublespace}
\begin{lem}
\begin{doublespace}
\label{prop:Truncated_N_The-Bhattacharyya-coefficient}The Bhattacharyya
distance, when we have truncated normal distributions $p,q$ that
do not overlap, is zero and when they overlap, it is given by 
\begin{eqnarray*}
D_{BC-TN}(p,q) & = & \frac{1}{4}\left(\frac{(\mu_{p}-\mu_{q})^{2}}{\sigma_{p}^{2}+\sigma_{q}^{2}}\right)+\frac{1}{4}\ln\left(\frac{1}{4}\left(\frac{\sigma_{p}^{2}}{\sigma_{q}^{2}}+\frac{\sigma_{q}^{2}}{\sigma_{p}^{2}}+2\right)\right)\\
 &  & +\frac{1}{2}\ln\left[\Phi\left(\frac{b-\mu_{p}}{\sigma_{p}}\right)-\Phi\left(\frac{a-\mu_{p}}{\sigma_{p}}\right)\right]+\frac{1}{2}\ln\left[\Phi\left(\frac{d-\mu_{q}}{\sigma_{q}}\right)-\Phi\left(\frac{c-\mu_{q}}{\sigma_{q}}\right)\right]\\
 &  & -\ln\left\{ \Phi\left[\frac{u-\nu}{\varsigma}\right]-\Phi\left[\frac{l-\nu}{\varsigma}\right]\right\} 
\end{eqnarray*}
Here,
\[
p\sim N\left(\mu_{p},\sigma_{p}^{2},a,b\right)\;;\;q\sim N\left(\mu_{q},\sigma_{q}^{2},c,d\right)
\]
\[
l=\min\left(a,c\right)\;;\;u=\min\left(b,d\right)
\]
\[
\nu=\frac{\left(\mu_{p}\sigma_{q}^{2}+\mu_{q}\sigma_{p}^{2}\right)}{\left(\sigma_{p}^{2}+\sigma_{q}^{2}\right)}\;;\;\varsigma=\sqrt{\frac{2\sigma_{p}^{2}\sigma_{q}^{2}}{\left(\sigma_{p}^{2}+\sigma_{q}^{2}\right)}}
\]
\end{doublespace}

\end{lem}
\begin{doublespace}

\subsection{Truncated Multivariate Normal Distribution}
\end{doublespace}

\begin{doublespace}
Similarly, a truncated multivariate normal distribution $\boldsymbol{X}$
has the density function, 
\[
f_{\mathbf{X}}\left(x_{1},\ldots,x_{k}\mid\boldsymbol{\mu_{p}},\,\boldsymbol{\Sigma_{p}},\,\boldsymbol{a},\,\boldsymbol{b}\right)=\frac{\exp\left(-\frac{1}{2}\left({\mathbf{x}}-{\boldsymbol{\mu_{p}}})^{\mathrm{T}}\right){\boldsymbol{\Sigma_{p}}}^{-1}\left({\mathbf{x}}-{\boldsymbol{\mu_{p}}}\right)\right)}{\int_{\boldsymbol{a}}^{\boldsymbol{b}}\exp\left(-\frac{1}{2}\left({\mathbf{x}}-{\boldsymbol{\mu_{p}}})^{\mathrm{T}}\right){\boldsymbol{\Sigma_{p}}}^{-1}\left({\mathbf{x}}-{\boldsymbol{\mu_{p}}}\right)\right)d\boldsymbol{x};\;\boldsymbol{x}\in\boldsymbol{R}_{\boldsymbol{a}\leq\boldsymbol{x}\leq\boldsymbol{b}}^{k}}
\]
Here, $\boldsymbol{\mu_{p}}$ is the mean vector and $\boldsymbol{\Sigma_{p}}$
is the symmetric positive definite covariance matrix of the $\boldsymbol{p}$
distribution and the integral is a $k$ dimensional integral with
lower and upper bounds given by the vectors $\left(\boldsymbol{a},\boldsymbol{b}\right)$
and $\boldsymbol{\boldsymbol{x}\in\boldsymbol{R}_{\boldsymbol{a}\leq\boldsymbol{x}\leq\boldsymbol{b}}^{k}}$
. 
\end{doublespace}
\begin{lem}
\begin{doublespace}
\label{prop:Truncated_MN_The-Bhattacharyya-coefficient}The Bhattacharyya
coefficient when we have truncated multivariate normal distributions
$\boldsymbol{p},\boldsymbol{q}$ and all the $k$ dimensions have
some overlap, is given by 
\begin{eqnarray*}
D_{BC-TMN}\left(\boldsymbol{p},\boldsymbol{q}\right) & = & \frac{1}{8}(\boldsymbol{\mu_{p}}-\boldsymbol{\mu_{q}})^{T}\boldsymbol{\Sigma}^{-1}(\boldsymbol{\mu_{p}}-\boldsymbol{\mu_{q}})+\frac{1}{2}\ln\,\left(\frac{\det\boldsymbol{\Sigma}}{\sqrt{\det\boldsymbol{\Sigma_{p}}\,\det\boldsymbol{\Sigma_{q}}}}\right)\\
 &  & +\frac{1}{2}\ln\left[\frac{1}{\sqrt{(2\pi)^{k}\left(|\boldsymbol{\Sigma_{p}}|\right)}}\int_{\boldsymbol{a}}^{\boldsymbol{b}}\exp\left(-\frac{1}{2}\left({\mathbf{x}}-{\boldsymbol{\mu_{p}}})^{\mathrm{T}}\right){\boldsymbol{\Sigma_{p}}}^{-1}\left({\mathbf{x}}-{\boldsymbol{\mu_{p}}}\right)\right)d\boldsymbol{x};\;\boldsymbol{x}\in\boldsymbol{R}_{\boldsymbol{a}\leq\boldsymbol{x}\leq\boldsymbol{b}}^{k}\right]\\
 &  & +\frac{1}{2}\ln\left[\frac{1}{\sqrt{(2\pi)^{k}\left(|\boldsymbol{\Sigma_{q}}|\right)}}\int_{\boldsymbol{c}}^{\boldsymbol{d}}\exp\left(-\frac{1}{2}\left({\mathbf{x}}-{\boldsymbol{\mu_{\boldsymbol{q}}}})^{\mathrm{T}}\right){\boldsymbol{\Sigma_{\boldsymbol{q}}}}^{-1}\left({\mathbf{x}}-{\boldsymbol{\mu_{\boldsymbol{q}}}}\right)\right)d\boldsymbol{x};\;\boldsymbol{x}\in\boldsymbol{R}_{\boldsymbol{\boldsymbol{c}}\leq\boldsymbol{x}\leq\boldsymbol{d}}^{k}\right]\\
 &  & -\ln\left[\frac{1}{\sqrt{(2\pi)^{k}\det\left({\boldsymbol{\Sigma_{p}}}{\boldsymbol{\Sigma}}^{-1}{\boldsymbol{\Sigma_{q}}}\right)}}\right.\\
 &  & \left.\int_{\boldsymbol{l}}^{\boldsymbol{u}}\exp\left(-\frac{1}{2}\left\{ \left({\mathbf{x}-\mathbf{m}}\right)^{\mathrm{T}}\left({\boldsymbol{\Sigma_{q}}}^{-1}\left[{\boldsymbol{\Sigma}}\right]{\boldsymbol{\Sigma_{p}}}^{-1}\right)\left({\mathbf{x}-\mathbf{m}}\right)\right\} \right)d\boldsymbol{x};\;\boldsymbol{x}\in\boldsymbol{R}_{\boldsymbol{\min\left(a,\boldsymbol{c}\right)}\leq\boldsymbol{x}\leq\boldsymbol{\min\left(b,d\right)}}^{k}\vphantom{\frac{\sqrt{(2\pi)^{k}\det\left({\boldsymbol{\Sigma_{p}}}^{-1}\right)}}{\sqrt{(2\pi)^{k}\det\left({\boldsymbol{\Sigma_{p}}}^{-1}\right)}}}\right]
\end{eqnarray*}
Here,
\[
\boldsymbol{p}\sim N\left(\boldsymbol{\mu_{p}},\,\boldsymbol{\Sigma_{p}},\,\boldsymbol{a},\,\boldsymbol{b}\right)
\]
\[
\boldsymbol{q}\sim N\left(\boldsymbol{\mu_{\boldsymbol{q}}},\,\boldsymbol{\Sigma_{\boldsymbol{q}}},\,\boldsymbol{c},\,\boldsymbol{d}\right)
\]
\[
\boldsymbol{u=\min\left(b,d\right)}\;;\;\boldsymbol{l=\min\left(a,c\right)}
\]
\[
{\mathbf{m}}=\left[\left({\boldsymbol{\mu_{p}}}^{\mathrm{T}}{\boldsymbol{\Sigma_{p}}}^{-1}+{\boldsymbol{\mu_{q}}}^{\mathrm{T}}{\boldsymbol{\Sigma_{q}}}^{-1}\right)\left({\boldsymbol{\Sigma_{p}}}^{-1}+{\boldsymbol{\Sigma_{q}}}^{-1}\right)^{-1}\right]^{\mathrm{T}}
\]
\[
\boldsymbol{\Sigma}=\frac{\boldsymbol{\Sigma_{p}}+\boldsymbol{\Sigma_{q}}}{2}
\]
\end{doublespace}

\end{lem}
\begin{doublespace}

\subsection{Covariance and Distance}
\end{doublespace}

\begin{doublespace}
The following is a general extension to Stein's lemma (Stein 1973,
1981; Rubinstein 1973, 1976) that does not require normality, involving
the covariance between a random variable and a function of another
random variable. Kattumannil (2009) extends the Stein lemma by relaxing
the requirement of normality. (Teerapabolarn 2013) is a further extension
of this normality relaxation to discrete distributions. Another useful
reference, (Kimeldorf and Sampson 1973), provides a class of inequalities
between the covariance of two random variables and the variance of
a function of the two random variables.
\end{doublespace}
\begin{lem}
\begin{doublespace}
\label{prop:Distance-Covariance-Relationship}The following equations
govern the relationship between the Bhattacharyya distance, $\rho\left(f_{X},f_{Y}\right)$,
and the covariance between any two distributions with joint density
function, $f_{XY}\left(t,u\right)$, means, $\mu_{X}\text{ and }\mu_{Y}$
and density functions $f_{X}\left(t\right)\text{ and }f_{Y}\left(t\right)$,
\[
\text{Cov}\left[c\left(X\right),Y\right]=\text{Cov}\left(X,Y\right)-E\left[\sqrt{\frac{f_{Y}\left(t\right)}{f_{X}\left(t\right)}}Y\right]+\mu_{Y}\:\rho\left(f_{X},f_{Y}\right)
\]
\[
\text{Cov}\left(X,Y\right)+\mu_{Y}\:\rho\left(f_{X},f_{Y}\right)=E\left[c'\left(X\right)\:g\left(X,Y\right)\right]+E\left[\sqrt{\frac{f_{Y}\left(t\right)}{f_{X}\left(t\right)}}Y\right]
\]
Here, 
\[
c\left(t\right)=t-\sqrt{\frac{f_{Y}\left(t\right)}{f_{X}\left(t\right)}}
\]
and $g\left(t,u\right)$ is a non-vanishing function such that,

\[
\frac{f_{XY}'\left(t,u\right)}{f_{XY}\left(t,u\right)}=-\frac{g'\left(t,u\right)}{g\left(t,u\right)}+\frac{\left[\mu_{Y}-u\right]}{g\left(t,u\right)}\quad,\quad t,u\in(a,b)
\]
\end{doublespace}
\end{lem}

\end{document}